\documentclass{appolb}

\usepackage[numbers,sort&compress]{natbib}
\usepackage{graphicx}
\usepackage{bm}
\usepackage{amsmath}
\usepackage{subfigure}
\usepackage[usenames,dvipsnames]{color}
\usepackage[colorlinks=false,linktocpage=true,linkcolor=blue,citecolor=blue]{hyperref}

\newcommand{\be}{\begin{equation}}
\newcommand{\ee}{\end{equation}}
\newcommand{\ba}{\begin{eqnarray}}
\newcommand{\ea}{\end{eqnarray}}
\newcommand{\nn}{\nonumber}

\def\lsim{\mathrel{\rlap{\lower4pt\hbox{$\sim$}}
    \raise1pt\hbox{$<$}}}                
\def\gsim{\mathrel{\rlap{\lower4pt\hbox{$\sim$}}
    \raise1pt\hbox{$>$}}}                

\newcommand{\mysubsubsection}[1]{\vspace{2mm}\noindent{\em #1}\vspace{2mm}}

\begin{document}

\eqsec  

\title{Anisotropic Hydrodynamics:  Three lectures%
\thanks{Presented at the LIV Cracow School of Theoretical Physics, Zakopane, Poland, June 2014.}}
\author{Michael Strickland%
\address{Department of Physics, Kent State University, Kent, OH 44242 United States}}
\maketitle

\begin{abstract}
Anisotropic hydrodynamics is a non-perturbative reorganization of relativistic hydrodynamics that takes into account the large momentum-space anisotropies generated in ultrarelativistic heavy-ion collisions.  As a result, it allows one to extend the regime of applicability of hydrodynamic treatments to situations that can be quite far from isotropic thermal equilibrium.  In this paper, I review the material presented in a series of three introductory lectures.  I review the derivation of ideal and second-order viscous hydrodynamics from kinetic theory.  I then show how to extend the methods used to a system that can be highly anisotropic in local-rest-frame momenta.  I close by discussing recent work on this topic and then present an outlook to the future.
\end{abstract}

  
\section{Introduction}

The use of relativistic viscous hydrodynamics to model the spatiotemporal evolution of the quark-gluon plasma (QGP) generated in ultrarelativistic heavy-ion collisions (URHICs) is now widespread.  Results from such simulations are in quite good agreement with experimental data available from the Relativistic Heavy Ion collider (RHIC) at Brookhaven National Lab and the Large Hadron Collider at the European Center for Nuclear Research (CERN) \cite{Huovinen:2001cy,Hirano:2002ds,Kolb:2003dz,Muronga:2001zk,Muronga:2003ta,Muronga:2004sf,Heinz:2005bw,Baier:2006um,Romatschke:2007mq,Baier:2007ix,Dusling:2007gi,Luzum:2008cw,Song:2008hj,Heinz:2009xj,El:2009vj,PeraltaRamos:2009kg,El:2010mt,PeraltaRamos:2010je,Denicol:2010tr,Denicol:2010xn,Schenke:2010rr,Schenke:2011tv,Bozek:2011wa,Niemi:2011ix,Niemi:2012ry,Bozek:2012qs,Denicol:2012cn,Denicol:2012es,PeraltaRamos:2012xk,Jaiswal:2013npa,Jaiswal:2013vta,Calzetta:2014hra,Denicol:2014vaa,Denicol:2014mca,Jaiswal:2014isa}.  Despite this success, there are still some fundamental issues to be addressed in the context of relativistic hydrodynamics applied to heavy ion collisions.  One of these issues stems from the fact that the traditional derivation of the dynamical equations of viscous hydrodynamics relies on a linearization around an isotropic equilibrium distribution function.  In recent years, we have come to understand that the QGP, as generated in URHICs is not momentum space isotropic.  In fact, at very early times after the initial nuclear impact one finds very large pressure anisotropies, i.e. ${\cal P}_L/{\cal P}_T \lsim 0.3$ in the center of the fireball if one uses shear viscosities that are consistent with experimental observations.  As one move out from the center of the fireball to the colder regions of the plasma, the level of plasma anisotropy increases.  Such large pressure anisotropies are an indicator of large viscous corrections to the assumed starting point of ideal hydrodynamics.  In addition, one finds that using traditional linearized viscous hydrodynamical treatments, that there always exist regions of phase space in which the one-particle distribution function is negative.  The size of these unphysical regions increases as one considers early times or colder regions of the plasma.  Since the one-particle distribution function is used as input to phenomenological calculations of other plasma signatures, such as photon and dilepton production/flow, quarkonium suppression, and freeze-out, this can potentially lead to inaccuracies in model calculations.\footnote{For phenomenological calculations of these signatures and related theoretical developments, I refer the reader to Refs.~\cite{Mrowczynski:1993qm,Mrowczynski:2000ed,Randrup:2003cw,Romatschke:2003ms,Romatschke:2003vc,Romatschke:2004jh,Romatschke:2004au,Mrowczynski:2004kv,Schenke:2006fz,Schenke:2006yp,Chatterjee:2007xk,Mauricio:2007vz,Dumitru:2007rp,Dumitru:2007hy,Martinez:2008di,Dusling:2008xj,Dusling:2008nt,Bhattacharya:2008up,Bhattacharya:2008mv,Rebhan:2008uj,Dumitru:2009ni,Burnier:2009yu,Dumitru:2009fy,Philipsen:2009wg,Bhattacharya:2009sb,Dusling:2009bc,Bhattacharya:2010sq,Margotta:2011ta,Strickland:2011mw,Strickland:2011aa,Mandal:2011xn,Mandal:2011jx,Dion:2011pp,Carrington:2011uj,Florkowski:2012ax,Strickland:2012cq,Attems:2012js,Mandala:2013xma,Shen:2013cca,Shen:2013vja,Ryblewski:2013eja,Vujanovic:2013jpa,Jahnke:2013rca,Mandal:2013jla,Fadafan:2013bva,Shen:2014cga,Shen:2014nfa,Rougemont:2014efa,Carrington:2014bla,Kumar:2014fta,Ali-Akbari:2014xea,Cheng:2014qia}.}

Because of the aforementioned problems, there was motivation to create an alternative framework for describing dissipative dynamics that could more accurately describe the dynamics of highly momentum-space anisotropic and, hence, far-from-equilibrium systems.  One method that has been proven to be quite successful is the framework of {\em anisotropic hydrodynamics} \cite{Martinez:2010sc,Florkowski:2010cf,Ryblewski:2010bs,Martinez:2010sd,Ryblewski:2011aq,Florkowski:2011jg,Martinez:2012tu,Ryblewski:2012rr,Florkowski:2012as,Florkowski:2013uqa,Ryblewski:2013jsa,Bazow:2013ifa,Tinti:2013vba,Florkowski:2014bba,Florkowski:2014txa,Nopoush:2014pfa,Denicol:2014mca}.  In this framework one allows the leading-order (LO) one-particle distribution function to be momentum-space anisotropic.  The most important anisotropies are of spheroidal form, i.e. $T^{xx}_{\rm LO} = T^{yy}_{\rm LO} \neq T^{zz}_{\rm LO}$ in the local rest frame (LRF) \cite{Romatschke:2003ms,Martinez:2010sc,Florkowski:2010cf}, however, it is possible to start with a more general ellipsoidal momentum-space anisotropy, i.e. $T^{xx}_{\rm LO} \neq T^{yy}_{\rm LO} \neq T^{zz}_{\rm LO}$ in the LRF \cite{Tinti:2013vba,Nopoush:2014pfa,Denicol:2014mca}.  With either prescription, the starting point for the derivation of the anisotropic hydrodynamics equations is to assume that one can express the one-particle distribution function in the form
\begin{equation}
   f(x,p) = 
   \underbrace{f_{\rm iso}\!\left(\frac{\sqrt{p^\mu \Xi_{\mu\nu}(x)p^\nu}}{\Lambda(x)}, \frac{\mu(x)}{\Lambda(x)}\right)}_{f_{\rm aniso}(x,p)}
   + \, \delta\!\tilde{f}(x,p) \, ,
\label{eq:ahexp}
\end{equation}
where $\Xi_{\mu\nu}$ is a second-rank tensor that measures the amount of momentum-space anisotropy and $\Lambda$ is a temperature-like scale which can be identified with the true temperature of the system in the isotropic equilibrium limit. $\mu(x)$ is the effective chemical potential of the particles. Traditionally, LO anisotropic hydrodynamics (aHydro) is based on an azimuthally symmetric (spheroidal) ansatz for $\Xi_{\mu\nu}(x)$ \cite{Romatschke:2003ms,Martinez:2010sc,Florkowski:2010cf} involving a single anisotropy parameter $\xi$ such that $p^\mu \Xi_{\mu\nu}(x)p^\nu$ reduces to ${\bf p}^2 + \xi(x) p_L^2$ in the LRF. The dynamical equations of LO spheroidal aHydro were originally derived from kinetic theory by taking $f(x,p) = f_{\rm aniso}(x,p)$ (i.e. by ignoring the correction $\delta\!\tilde{f}$ in Eq.~(\ref{eq:ahexp})), and using the zeroth and first moments of the Boltzmann equation in the relaxation time approximation \cite{Martinez:2010sc,Florkowski:2010cf,Florkowski:2011jg,Martinez:2012tu}.

Using a spheroidal form at LO is motivated by the fact that it accounts for the most important anisotropic corrections to the one-particle distribution non-perturbatively.  In addition, it benefits from the following properties
\begin{itemize}
\itemsep3pt \parskip0pt \parsep0pt
\item It gives the ideal hydro limit when $\xi \rightarrow 0$ ($\Lambda \rightarrow T$) which corresponds to the limit $\eta/{\cal S} \rightarrow 0$.
\item It gives the longitudinal free-streaming limit for a transversely homogeneous system undergoing boost-invariant longitudinal expansion (0+1d expansion), which corresponds to the limit $\eta/{\cal S} \rightarrow \infty$.  This is an extreme case where the system develops the maximal degree of momentum-space anisotropy and it is exactly described by a spheroidal form.
\item Since $f_{\rm iso} \geq 0$, the one-particle distribution function and pressures are all greater than or equal to zero.  This is not guaranteed in linearized viscous hydrodynamics.
\item It can be shown that the aHydro formalism reduces to linearized second-order viscous hydrodynamics in the limit of small anisotropies.  This was originally shown in the context of 0+1d expansion in Ref.~\cite{Martinez:2010sc}, but can be shown to hold also in case of fully 3+1d dynamics \cite{Tinti:forth}
\end{itemize}

In the last year, the corrections due to $\delta\!\tilde{f}$ in (\ref{eq:ahexp}) were included in a next-leading-order (NLO) treatment of anisotropic hydrodynamics~\cite{Bazow:2013ifa}.  At NLO, dissipative effects due to the spheroidally deformed LO term are treated non-perturbatively, while the non-spheroidal corrections $\delta\!\tilde{f}$ are treated perturbatively.  Another interesting recent development has been to generalize the RS form from spheroidal to ellipsoidal form at LO \cite{Tinti:2013vba}, at least for the case of a system which possesses cylindrical symmetry in space.  This development offers some promise to treat all diagonal components of the energy-momentum tensor non-perturbatively, while treating only the off-diagonal components perturbatively. 

The goal of the anisotropic hydrodynamics program is to create a dissipative hydrodynamics framework that more accurately describes
\begin{itemize}
\itemsep3pt \parskip0pt \parsep0pt
\item Early time dynamics of the QGP created in heavy-ion collisions.
\item Dynamics near the transverse edges of the nuclear overlap region
\item Temperature-dependent (and potentially large) $\eta/{\cal S}$
\end{itemize}
In the following three sections, I will attempt to summarize the content presented in the three lectures I gave in the LIV Cracow School of Theoretical Physics.  The first lecture discussed motivation and evidence for large momentum-space anisotropes and then proceeded to the derivation of ideal hydrodynamics from kinetic theory.  The second lecture discussed linearized viscous hydrodynamics and anisotropic hydrodynamics.  The third lecture discussed recent advances which have included the development of exact solutions to the Boltzmann equation in the relaxation time approximation and comparisons of the results obtained in various hydrodynamics frameworks to these exact solutions.  

In this manuscript, I use the particle-physics Minkowski-space metric convention $g^{\mu\nu} = {\rm diag}(+1,-1,-1,-1)$ and natural units with $\hbar = c = k_B = 1$.  With this metric convention, the flow velocity $u^\mu$ is normalized as  $u_\mu u^\mu=1$.   Milne coordinates in Minkowski space are defined by $x^{\mu }=(\tau ,x,y,\varsigma)$, with longitudinal proper time $\tau =\sqrt{t^2-z^2}$, spacetime rapidity $\varsigma ={\rm arctanh}(z/t)$,  and metric $ds^2=d\tau^2-dx^2-dy^2-\tau^2d\varsigma^2$. In some places, I denote the scalar product between two 4-vectors with a dot, i.e. $A_{\mu }B^{\mu }\equiv A\cdot B$.

\section{Lecture 1}

In lecture 1, I provide motivation for the study of anisotropic hydrodynamics and then turn to the derivation of ideal hydrodynamics from kinetic theory.

\subsection{Motivation}

\begin{figure}[t]
\centerline{\includegraphics[width=0.85\linewidth]{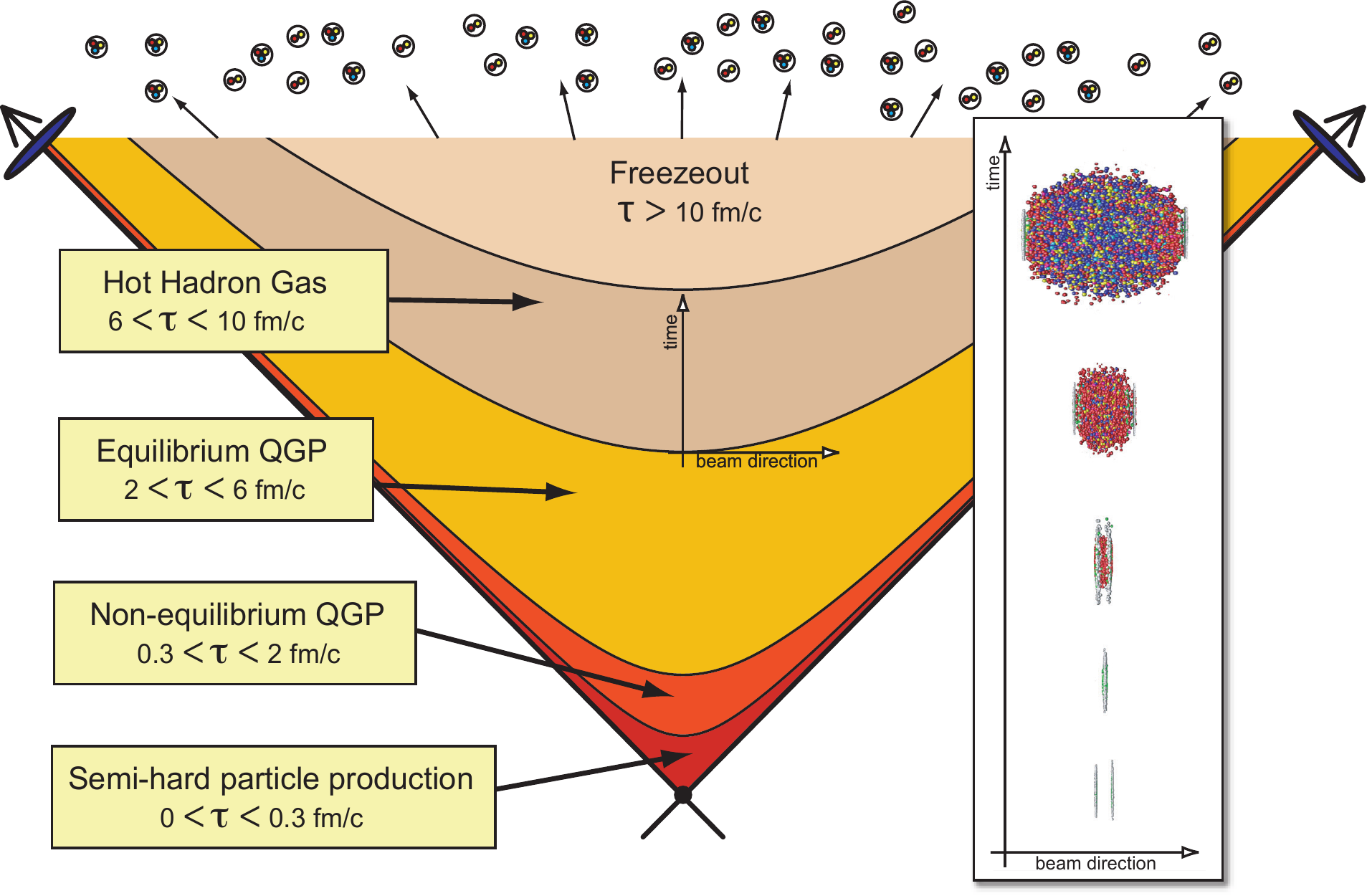}}
\caption{(Color online) A cartoon depicting the space-time history of the QGP as generated in a heavy ion collision at LHC energies.  The overlay on the right shows the lab-frame evolution.}
\label{fig:spacetimehistory}
\end{figure}

If we visualize a 2d slice of the space-time history of the quark-gluon plasma as generated in a URHIC at the LHC, it would look something like the cartoon shown in Fig.~\ref{fig:spacetimehistory}.  The timescales shown should be taken as rough guidelines rather than hard and fast numbers, but these estimates are reasonable if one considers the evolution of the central region of the QGP fireball.  As this figure shows, at the earliest time after the initial nuclear impact, QGP evolution is dominated by hard particle production processes which, in the high-energy limit, are describable in terms of the color-glass-condensate (CGC)/glasma \cite{McLerran:1993ni,McLerran:1993ka,Iancu:2003xm,Lappi:2006fp,Gelis:2009wh,Gelis:2010nm}.  During this time period, one expects extremely large deviations from isotropic thermal equilibrium.  In fact, due to the coherent fields that exist in the CGC initial state, the initial longitudinal pressure is expected to be negative.  After a few multiples of the inverse gluon saturation scale, $Q_s^{-1} \sim 0.1{-}0.2$ fm/c, however, this negative pressure goes away and one finds a positive, but very small, longitudinal pressure \cite{Lappi:2006fp}.  In the past, the large amount of momentum-space anisotropy present in CGC-like initial conditions has been used as argument against the application of perturbative QCD (pQCD) to heavy-ion collisions since it was thought that one could not match onto hydrodynamical solutions in this case.  However, it turns out that strong-coupling approaches and viscous hydrodynamics itself also predict large momentum-space anisotropies at early times.  The existence of large momentum-space anisotropies in the QGP seems to be very much model-independent.

Looking again at Fig.~\ref{fig:spacetimehistory}, we see that, after the initial period of hard particle production, there is a pre-equilibrium period that may extend for as long as 2 fm/c.  In the past, it has been claimed that the pre-equilibrium period can only exist for up to 1 fm/c and that, after that, the QGP becomes isotropic; however, we now understand that viscous hydrodynamics itself shows large corrections to ideal isotropic behavior even at times as late as 2 fm/c.  After the pre-equilibrium period is over, one can begin to use linearized viscous hydrodynamics to describe the evolution of the QGP.\footnote{Of course, one can apply linearized viscous hydro prior to this time, but its reliability is less sure at early times.}  I emphasize, however, that these time scales are only appropriate for the description of the matter in the center of the fireball.  In a conformal system, the length of the pre-equilibrium stage scales like the inverse temperature.  Therefore, as one moves out of the center, towards the cooler transverse or longitudinal regions of the QGP, one expects much larger non-equilibrium corrections and a longer pre-equilibrium stage.

After the pre-equilibrium stage, we move into the hydrodynamic regime.  During this period, the expansion and cooling of the QGP can be described using the equations of linearized viscous hydrodynamics.  At late times, however, the system goes through a transition to hadronic degrees of freedom and eventually becomes too dilute to be reliably described by linearized viscous hydrodynamics once again.  The system subsequently ``freezes-out,'' first chemically and then kinetically, and finally, the produced hadrons free stream to the detectors, with an imprint of their former existence as a near-equilibrium QGP left on their spatial/momentum distributions and relative abundances.

\begin{figure}[t]
\centerline{\includegraphics[width=0.85\linewidth]{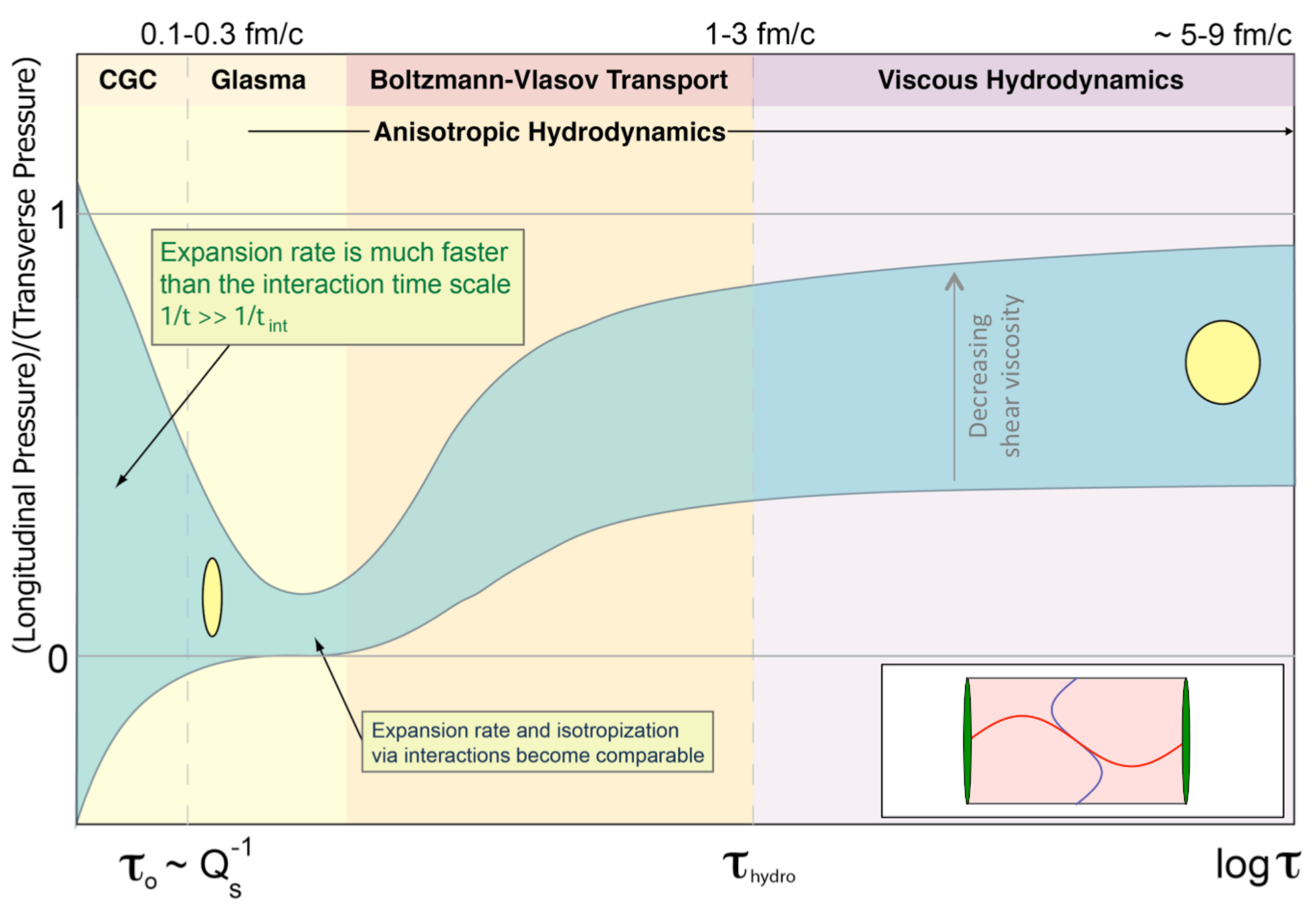}}
\caption{(Color online) A cartoon depicting the temporal evolution of the momentum-space anisotropy evolution expected to be generated in a heavy ion collision at LHC energies.  The inset yellow ellipses indicate the shape of the momentum-space distribution with the horizontal direction corresponding to the longitudinal (beamline) direction.  The inset in the lower right shows a snapshot of the receding nuclei, with the red wave indicating the stretching of a longitudinal mode and the blue wave indicating a pseudo-static transverse mode.}
\label{fig:anisotropyevol}
\end{figure}

Having discussed the general space-time picture of a heavy-ion collision, let's now discuss, in some more detail, the evolution of the level of pressure anisotropy expected.  In order to illustrate the pressure anisotropy expected at various stages of QGP evolution, in Fig.~\ref{fig:anisotropyevol}, I show a sketch of the proper-time evolution of the level of momentum-space anisotropy measured by the ratio of the longitudinal pressure, ${\cal P}_L$, and transverse pressures, ${\cal P}_T$.  The blue band shows a range for the possible level of momentum-space anisotropy.  At early times, the lower bound of this band illustrates the evolution expected in the CGC/glasma framework.  In addition to this early-time possibility, I also include another possibility, namely that the system starts in a prolate configuration (${\cal P}_L > {\cal P}_T$).  As this cartoon illustrates, however, one finds that the initial momentum-space anisotropy is not very important, since the rapid longitudinal expansion of the QGP pulls the system into a kind of universal attractor which results in an oblate (${\cal P}_L < {\cal P}_T$) momentum-space anisotropy at times on the order of a few tenths of fm/c.  The reason for this is that, in the high-energy limit and at early times, the longitudinal expansion scalar grows like $1/\tau$, while it takes some time ($\tau \sim R_T/c_s$) for collective effects to generate significant transverse expansion.  The effect of this is that, at early times, the QGP looks very much like tiny one-dimensionally expanding universe in which longitudinal momentum are strongly red-shifted while transverse momenta are largely unaffected.  As a result, the longitudinal pressure is strongly depleted relative to the transverse momentum.

After some time, however, the longitudinal expansion rate reduces and interactions among the QGP constituents are able to drive the system back towards isotropy.  However, as shown in Fig.~\ref{fig:anisotropyevol}, since the system is still longitudinally expanding, the interactions are never able to fully restore isotropy.  At late times, the degree of momentum-space anisotropy is set by the shear viscosity of the QGP as indicated in the figure.  If the shear viscosity to entropy density ratio ($\eta/{\cal S}$) is temperature-dependent, with large $\eta/{\cal S}$ at low- and high-temperatures, one can expect large non-equilibrium corrections at early and late times.  In addition, as already pointed out, if one moves to colder regions (not precisely in the center of the fireball), one finds that the level of momentum-space anisotropy increases and the length of time over which non-equilibrium corrections are important, becomes longer.

To sum up, one finds that there can be a sizable level of momentum-space anisotropy in the QGP.  Going beyond the cartoon level, it is possible to use both viscous hydrodynamics itself and also strong-coupling AdS/CFT approaches to try to reach some quantitative conclusions about the level of momentum-space anisotropy.  To arrive at some quantitative conclusions, let's consider first- and second-order viscous hydrodynamics for a system that is transversely homogeneous and boost invariant in the longitudinal direction, aka (0+1d)-dynamics.  In this case, first-order Navier Stokes (NS) viscous hydrodynamics predicts that the LRF shear correction to the ideal pressures is diagonal, with space-like components $\pi^{zz} = - 4\eta/3\tau = -2\pi^{xx} = - 2\pi^{yy}$, where $\eta$ is the shear viscosity and $\tau$ is the proper time.  In viscous hydrodynamics, the longitudinal pressure is given by ${\cal P}_L = P_{\rm eq} + \pi^{zz}$ and the transverse pressure by ${\cal P}_T =  P_{\rm eq} + \pi^{xx}$.  Assuming an ideal equation of state (EoS), the resulting ratio of the longitudinal pressure over the transverse pressure from first-order viscous hydrodynamics can be expressed as
\begin{equation}
\left(\frac{{\cal P}_L}{{\cal P}_T}\right)_{\rm \! NS} = \frac{ 3 \tau T - 16\bar\eta }{ 3 \tau T + 8\bar\eta } \, ,
\label{eq:aniso}
\end{equation}
where $\bar\eta \equiv \eta/{\cal S}$ with ${\cal S}$ being the entropy density.  Assuming RHIC-like initial conditions with $T_0 = 400$ MeV at $\tau_0 = 0.5$ fm/c and taking the conjectured lower bound $\bar\eta = 1/4\pi$ \cite{Policastro:2001yc}, one finds $\left({\cal P}_L/{\cal P}_T\right)_{\rm NS} \simeq 0.5$.  For LHC-like initial conditions with $T_0 = 600$ MeV at $\tau_0 = 0.25$ fm/c and once again taking $\bar\eta = 1/4\pi$ one finds $\left({\cal P}_L/{\cal P}_T\right)_{\rm NS} \simeq 0.35$.  This means that, even in the best case scenario of $\bar\eta = 1/4\pi$, viscous hydrodynamics itself predicts rather sizable momentum-space anisotropies.  For larger values of $\bar\eta$, one obtains even larger momentum-space anisotropies.  In addition, one can see from Eq.~(\ref{eq:aniso}) that, at fixed initial proper time, the level of momentum-space anisotropy increases as one lowers the temperature.

\begin{figure}[t]
\begin{center}
\includegraphics[width=0.49\linewidth]{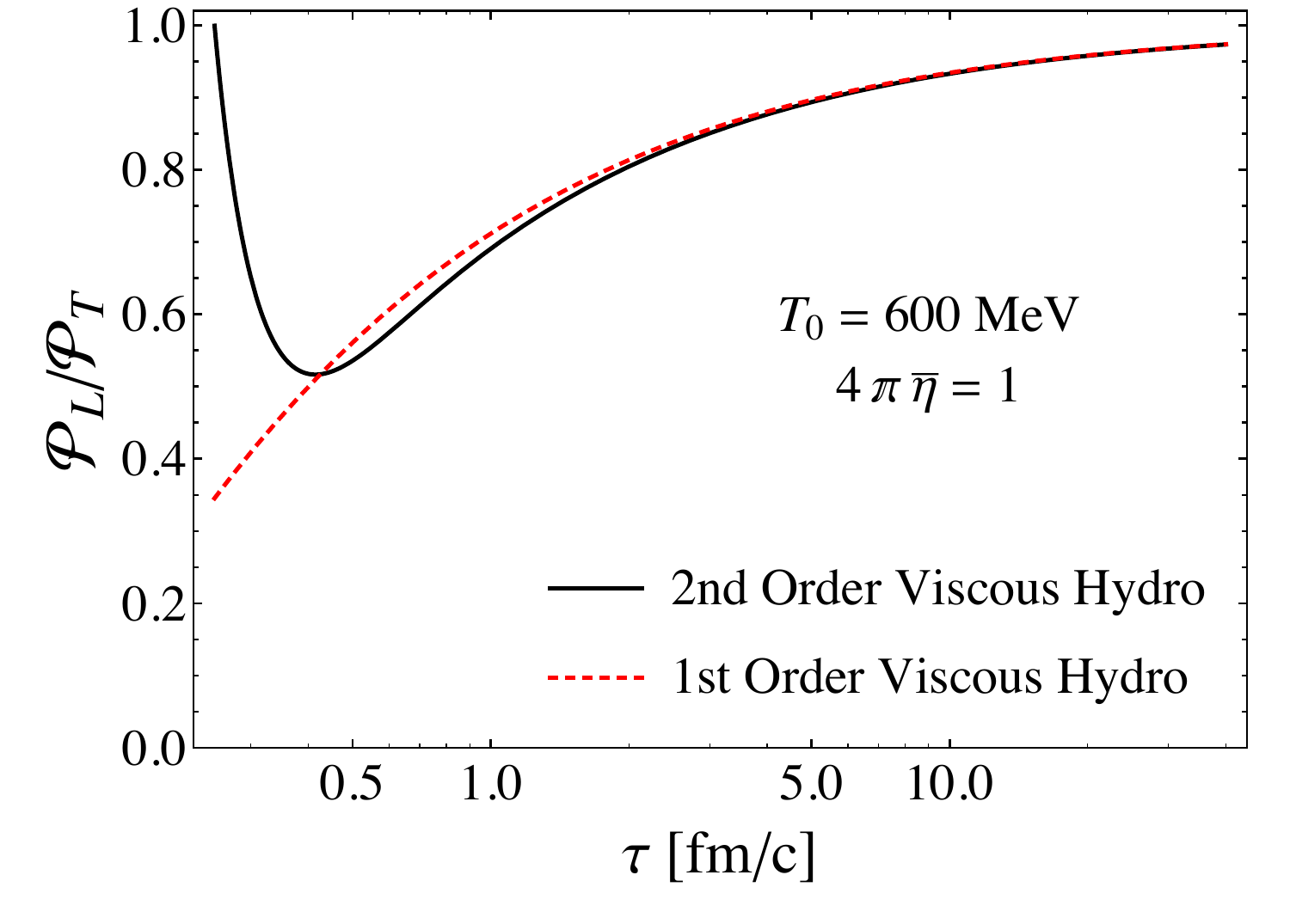}
\includegraphics[width=0.49\linewidth]{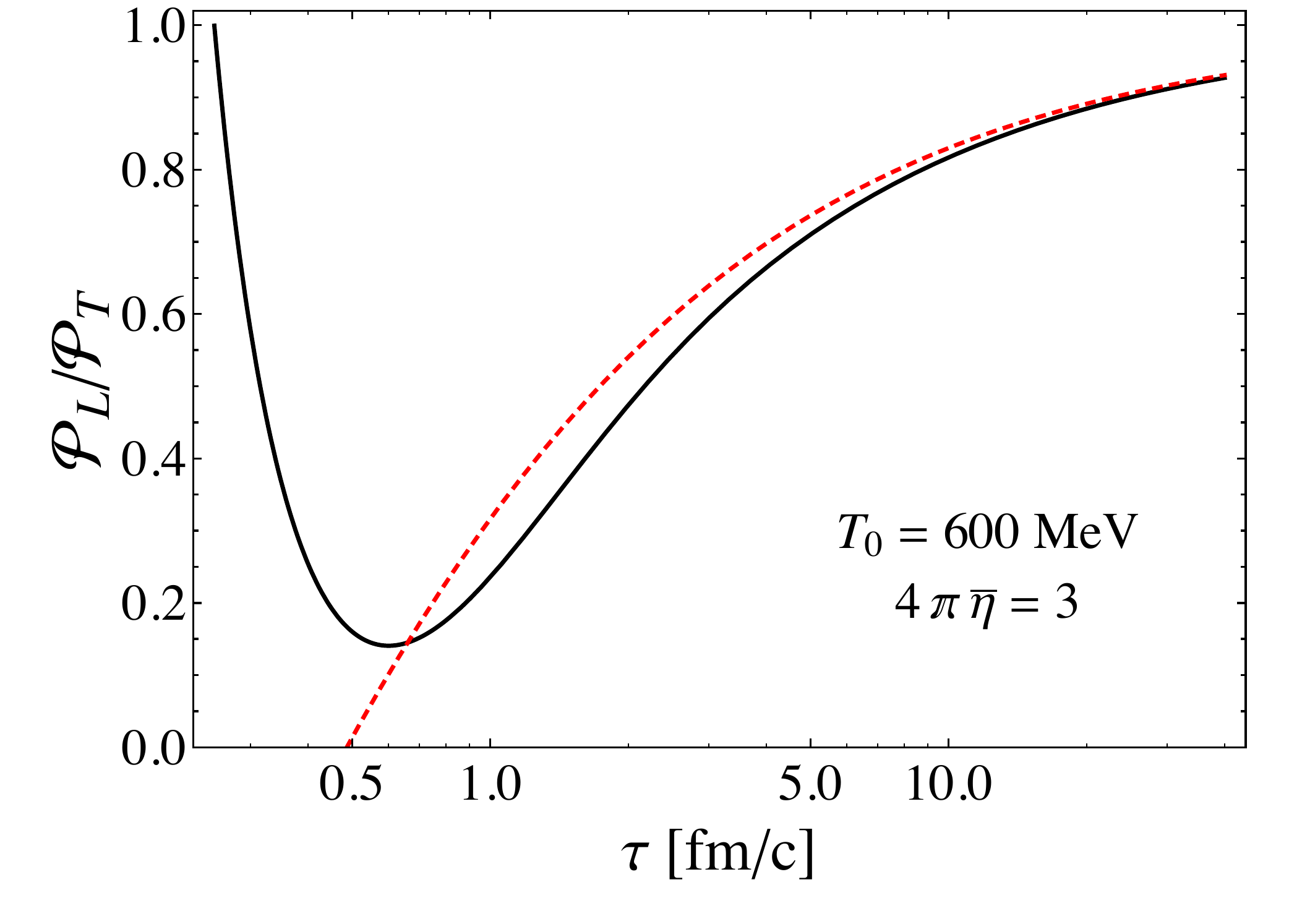}\\
\includegraphics[width=0.49\linewidth]{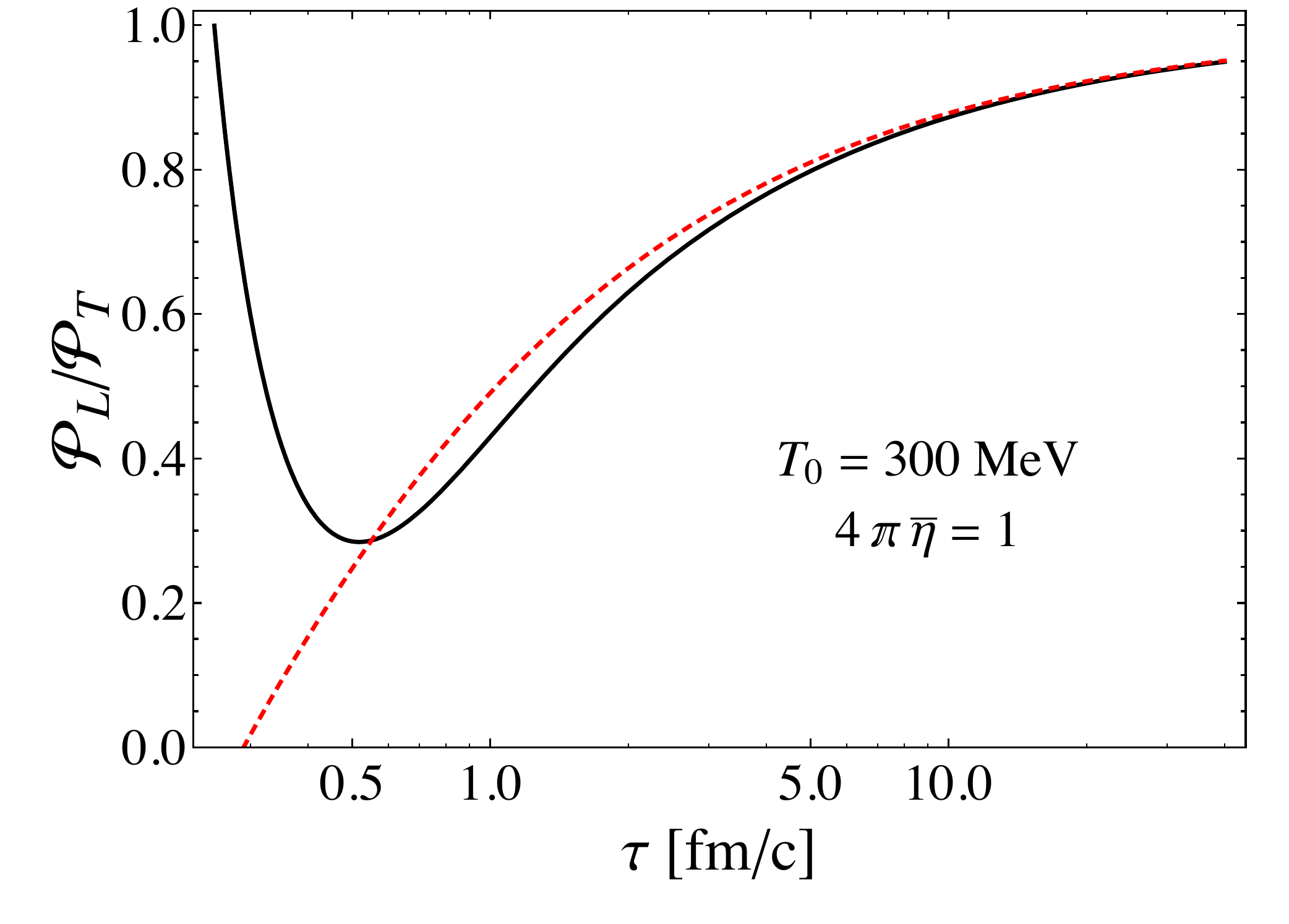}
\includegraphics[width=0.49\linewidth]{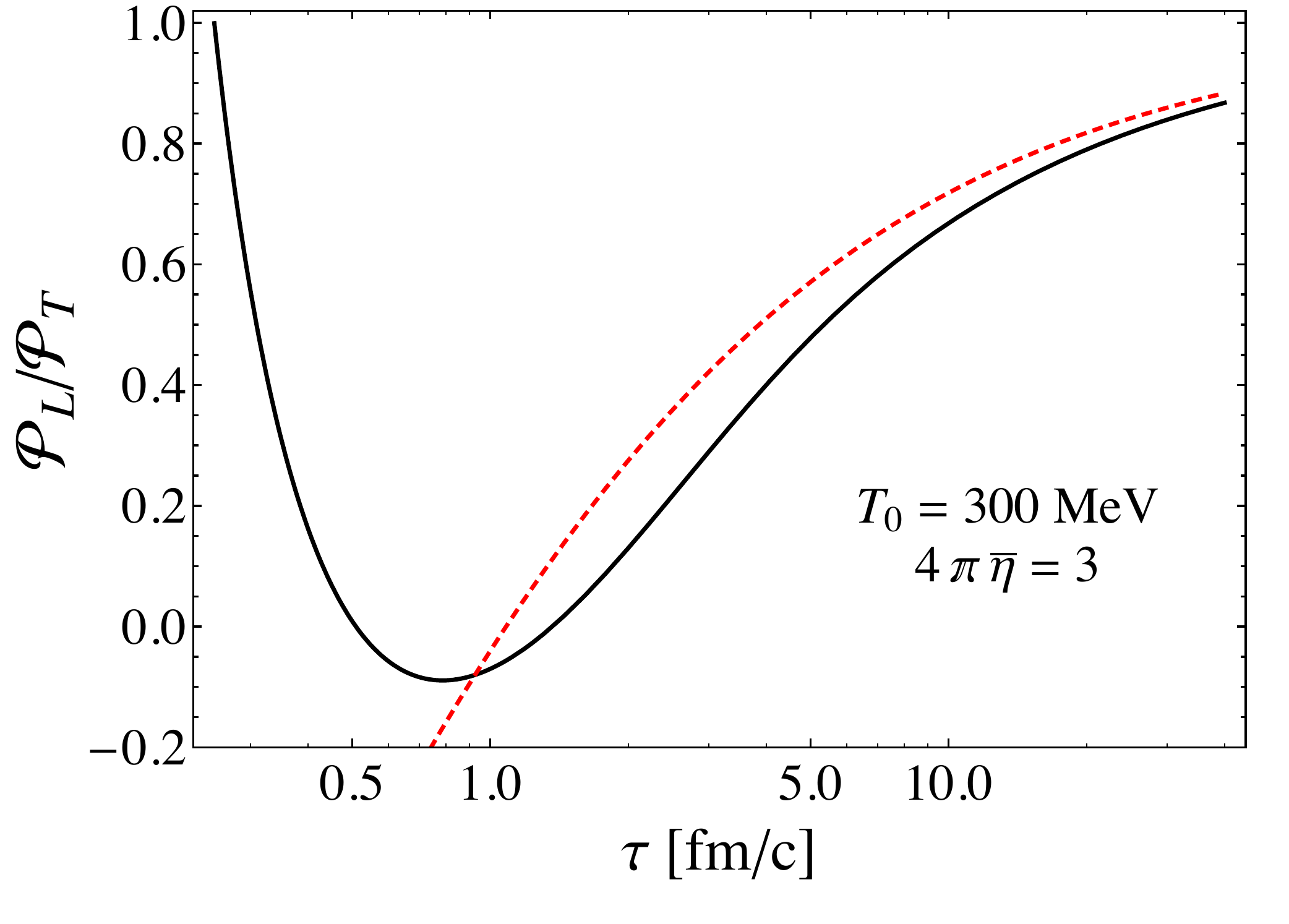}
\end{center}
\vspace{-7mm}
\caption{(Color online) The pressure anisotropy as a function of proper time assuming an initially isotropic system with $T_0 = 600$ MeV (top row) and $T_0 = 300$ MeV (bottom row) at $\tau_0 =$ 0.25 fm/c for $4\pi\bar\eta =$ 1 (left column) and 3 (right column).  The solid black line is the solution of the second order coupled differential equations and the red dashed line is the first-order Navier-Stokes solution.}
\label{fig:nscomp}
\end{figure}

Of course, since first-order relativistic viscous hydrodynamics is acausal, the analysis above is incomplete.  It does, however, provide intuitive guidance since the causal second-order version of the theory has the first-order solution as an attractor of the dynamics.  Because of this, one expects large momentum-space anisotropies to emerge within a few multiples of the shear relaxation time $\tau_\pi$.  In the strong-coupling limit of ${\cal N}=4$ SYM one finds $\tau_\pi = (2 - \log 2)/2 \pi T$ \cite{Baier:2007ix,Bhattacharyya:2008jc} which gives $\tau_\pi \sim 0.1$ fm/c and $\tau_\pi \sim 0.07$ fm/c for the RHIC- and LHC-like initial conditions stated above, respectively.  To demonstrate this quantitatively, in Fig.~\ref{fig:nscomp}, I plot the solution of the second order Israel-Stewart 0+1d viscous hydrodynamical equations assuming an isotropic initial condition and the NS solution together.  In the left column, I assumed $4\pi\bar\eta = 1$ and in the right column I assumed $4\pi\bar\eta = 3$ ($\bar\eta \simeq 0.24$) with $\tau_\pi = 2 (2 - \log 2) \bar\eta/T$ in both cases.  As can be seen from this figure, even if one starts with an isotropic initial condition, within a few multiples of the shear relaxation time one approaches the NS solution, overshoots it, and then approaches it from below.  The value of $\bar\eta$ in the right column is approximately the same as that extracted from recent fits to LHC collective flow data.  I note that if one further increases $\bar\eta$ or decreases the initial temperature, then one finds negative longitudinal pressures in second-order viscous hydrodynamics as well.  This can be seen in the lower right panel of Fig.~\ref{fig:nscomp}.  From this, one learns that the value of $\bar\eta$ extracted from LHC data \cite{Gale:2012rq} implies that the system may be highly momentum-space anisotropic.  In addition, from these figures we conclude that the momentum-space anisotropies persist throughout the evolution of the QGP.  

We can also ask what the expected degree of early-time momentum-space anisotropy within the context of the AdS/CFT framework is.  In this context, I mention the quite impressive work of two groups:  Heller et al. \cite{Heller:2011ju} and van der Schee et al. \cite{vanderSchee:2013pia}.  These two groups both simulated the dynamics of an expanding QGP using numerical general relativity (GR).  In the work of Heller et al.~\cite{Heller:2011ju}, they simulated the early time dynamics of a 0+1d system by numerically solving the GR equations in the bulk.  In the work of van der Schee et al.~\cite{vanderSchee:2013pia}, they performed similar numerical GR evolution, but in the case of a 1+1d radially symmetric system.  Both of these studies found early-time pressure anisotropies on the order of ${\cal P}_L/{\cal P}_T \sim 0.31$ or smaller.   Since these results were obtained in the context of the strong-coupling limit, which implies $4\pi\bar\eta=1$, the pressure anisotropy found is an upper bound on what to expect in reality.

Lastly, I would like to mention the role of plasma instabilities in the isotropization of a weakly-copuled QGP.  Currently, it is believed that the primary driving force for restoring isotropy in the gauge field sector are plasma instabilities such as the chromo-Weibel instability \cite{Strickland:2007fm}; however, practitioners have found that, even taking into account the unstable gauge field dynamics, the timescale for isotropization in classical Yang-Mills simulations is very long \cite{Romatschke:2006nk,Berges:2007re,Rebhan:2008uj,Attems:2012js}.  The recent work of Epelbaum and Gelis \cite{Gelis:2013rba} has included resummation of next-to-leading order (NLO) quantum loop corrections to the initial CGC fluctuations, and simulations in this framework find early-time pressure anisotropies on the order of $0.01\,$-$\,0.5$, depending on the assumed value of the strong coupling constant $g_s = 0.1\,$-$\,0.4$.  Other classical Yang-Mills simulations by Berges et al. \cite{Berges:2013eia,Berges:2013fga} also found persistent late-time momentum-space anisotropies, however, the ratio of ${\cal P}_L/{\cal P}_T$ they found was much smaller than obtained in the Epelbaum and Gelis simulations and, on top of that, they found that ${\cal P}_L/{\cal P}_T$ was a monotonically decreasing function of proper time at late proper times, suggesting a late-time anisotropic attractor.  In the context of hard-loop simulations of chromo-Weibel instability evolution, one finds rapid thermalization of the plasma in the sense that an anisotropic Boltzmann distribution of gluon modes is established within $\sim$ 1 fm/c; however, similar to other studies, one finds that large pressure anisotropies persist for at least $5\,$-$\,6$ fm/c \cite{Rebhan:2008uj,Attems:2012js}.

\subsection{Derivation of ideal hydrodynamics from kinetic theory}

I would now like to briefly review how one can obtain the ideal hydrodynamics equations of motion starting from kinetic theory.  I will restrict my considerations to the case that all chemical potential(s) are zero for simplicity.  The starting point for the derivation is the Boltzmann equation
\be
p^\mu \partial_\mu f(x,p) = - C[f(x,p)] \, ,
\ee
where $x^\mu = (t,{\bf x})$, $p^\mu = (E_p,{\bf p})$, $\partial_\mu = (\partial_t,-\nabla)$, and the functional $C$ is the collisional kernel which includes the effect of particle scattering.  To obtain the bulk equations of motion we take moments of the Boltzmann equation by multiplying the left and right by an integral operator of the form
\be
\hat{I}^{\nu_1 \nu_2 \cdots \nu_n} \; (\; \cdot \; ) = \int dP \prod_{i=1}^n p^{\nu_i} \; (\; \cdot \; ) \, ,
\ee
where $dP = d^3p/E(2\pi)^3$ is the Lorentz-invariant phase space measure.  Applying this operator at zeroth order to the Boltzmann equation, we obtain the zeroth moment of the Boltzmann equation
\ba
\int dP \, p^\mu \partial_\mu f &=& - \int dP \, C[f] 
\nn \\
\partial_\mu \left[\int dP \, p^\mu f \right] &=& - \int dP \, C[f] \, .
\ea
The quantity in square brackets above is simply the particle four-current $j^\mu = (\rho,{\bf j})$.  The right-hand side is the zeroth moment of the collision kernel.  To simplify forthcoming expressions, we can introduce a notation for the $n^{\rm th}$-moment of the collisional kernel, $C_i \equiv \int dP \, \prod_{i=1}^n p^{\mu_i} C[f]$, which allows us to write the zeroth-moment equation compactly as 
\be
\partial_\mu j^\mu = - C_0 \, .
\ee
For number-conserving theories, $C_0$ is zero and one obtains,
\be
\partial_\mu j^\mu = 0 \, ,
\label{eq:cont}
\ee
which is simply the relativistic continuity equation.  At this point, we can introduce a tensor basis for the particle current.  There are two four-vectors at our disposal $u^\mu$, which is the four-velocity of the local rest frame (fluid four-velocity), and $V^\mu$, which is transverse particle current ($u^\mu V_\mu=0$ by definition).  Note that $u^\mu$ is normalized such that $u^\mu u_\mu = 1$, which means that there are only three independent components of the four-velocity.  We can decompose the current into these two quantities 
\be
j^\mu = n u^\mu + V^\mu \, ,
\ee
where $n$ is the net charge density and $V^\mu = {\Delta^\mu}_\nu j^\nu$ is the diffusion current, where $\Delta^{\mu\nu} = g^{\mu\nu} - u^\mu u^\nu$ is the transverse projector which projects out the components of a four-vector that are orthogonal to $u^\mu$ and obeys ${\Delta^\mu}_\nu u^\nu =0$.  For ideal hydro, one can assume that particle flow and energy flow are same and, as a result, we can take $V^\mu \rightarrow 0$.  In this case, we simply have
\be
j^\mu = n u^\mu \, .
\ee
Plugging this into Eq.~(\ref{eq:cont}), we obtain
\ba
\partial_\mu ( n u^\mu )  &=& 0 
\nn \\
\underbrace{u^\mu \partial_\mu}_{\equiv D} n + n \underbrace{\partial_\mu u^\mu}_{\equiv \theta} &=& 0 \, ,
\ea
where $D$ is the comoving derivative and $\theta$ is the expansion scalar.  Using this notation, the zeroth moment for number-conserving theories can be written compactly as
\be
D n + n \theta = 0 \, .
\label{eq:zeromom}
\ee 

Next, we consider the first moment of the Boltzmann equation
\ba
\int dP \, p^\nu p^\mu \partial_\mu f &=& - C_1
\nn \\
\partial_\mu \underbrace{\left[ \int dP \, p^\mu p^\nu  f \right]}_{T^{\mu\nu}}  &=& - C_1 \, .
\ea
We recognize the quantity in square brackets as the energy-momentum tensor as indicated above.  For any theory that has a energy-momentum conserving collisional kernel (which is guaranteed in quantum field theory via the four-dimensional delta functions which enforce this), one has $C_1 = 0$.  We, therefore, obtain a simple result from the first moment
\be
\partial_\mu T^{\mu\nu} = 0 \, ,
\label{eq:firstmom}
\ee
which is the statement of energy-momentum conservation.  To proceed further, we need to establish a tensor basis for $T^{\mu\nu}$.  In ideal hydrodynamics, we assume that the system is isotropic in the local rest frame at all times.  As a result, there are only two structures that can appear in a rank-two tensor, namely $g^{\mu\nu}$ and $u^\mu u^\nu$.  Therefore, in ideal hydrodynamics we can always express
\be
T^{\mu\nu} = A u^\mu u^\nu + B g^{\mu\nu} \, ,
\label{eq:tmunuideal}
\ee
where $A$ and $B$ are unknown Lorentz-invariant coefficients.  In the local rest frame, one has $u^\mu_{\rm LRF} = (1,0,0,0)$ and, in ideal hydrodynamics, the energy-momentum tensor is diagonal with $T^{\mu\nu}_{\rm LRF} = {\rm diag}({\cal E},{\cal P},{\cal P},{\cal P})$. This allows us to fix $A$ and $B$ by evaluating $T^{\mu\nu}$ in the local rest frame.  Evaluating the $00$-component of $T^{\mu\nu}_{\rm LRF}$ gives ${\cal E} = A + B$.  Evaluating any of the three spacelike $ii$-components gives ${\cal P} = -B$.  Therefore, $A = {\cal E} + {\cal P}$ and $B = -{\cal P}$.  Plugging these results into (\ref{eq:tmunuideal}), we find
\ba
T^{\mu\nu} &=& ({\cal E} + {\cal P}) u^\mu u^\nu - {\cal P} g^{\mu\nu} 
\nn \\
&=& {\cal E} u^\mu u^\nu - {\cal P} \Delta^{\mu\nu} \, .
\label{eq:tmunuideal2}
\ea
Using this, we can turn Eq.~(\ref{eq:firstmom}) into a set four dynamical equations
\ba
\partial_\mu T^{\mu\nu} &=& \partial_\mu \left[ ({\cal E} + {\cal P}) u^\mu u^\nu - {\cal P} g^{\mu\nu} \right] = 0 
\nn \\
&=& u^\nu D ({\cal E} + {\cal P}) + ({\cal E} + {\cal P})(u^\nu \theta + D u^\nu) - \partial^\nu {\cal P} = 0 \, .
\ea
There are four equations above indexed by $\nu$.  In order to obtain four scalar equations, we can project these equations with $u_\nu$ and the transverse projector ${\Delta^\alpha}_\nu$.  Projecting with $u_\nu$ gives
\be
D ({\cal E} + {\cal P}) + ({\cal E} + {\cal P})(\theta + \hspace{-2mm} \underbrace{u_\nu D u^\nu}_{D(u^\nu u_\nu)=0} \hspace{-1.5mm}) - D {\cal P} = 0 \, .
\ee
Simplifying this expression, we obtain
\be
D {\cal E} + ({\cal E} + {\cal P}) \theta = 0 \, .
\ee
Projecting with ${\Delta^\alpha}_\nu$ gives
\be
({\cal E} + {\cal P}) {\Delta^\alpha}_\nu D u^\nu - \nabla^\alpha {\cal P} = 0 
\ee
where we have introduced the spatial gradient operator $\nabla^\alpha \equiv {\Delta^\alpha}_\nu \partial^\nu$.  To simplify this further, we can use $D({\Delta^\alpha}_\nu u^\nu) = 0$ to obtain ${\Delta^\alpha}_\nu D u^\nu = - u^\nu D{\Delta^\alpha}_\nu = u^\nu D(u^\alpha u_\nu) = D u^\alpha$ and we obtain
\be
({\cal E} + {\cal P}) D u^\alpha - \nabla^\alpha {\cal P} = 0 \, .
\ee
Where, above, it is understood that $\alpha$ is a spacelike index.  In what follows, I will simply replace $\alpha$ by $i$ to make this explicit.

Summarizing, we have obtained four equations from the first moment of the Boltzmann equation, one from the $u$ projection and three from the transverse projection
\ba
D {\cal E} + ({\cal E} + {\cal P}) \theta &=& 0 \, ,
\label{eq:firstmomidealeqs1}
\\
({\cal E} + {\cal P}) D u^i - \nabla^i {\cal P} &=& 0 \, .
\label{eq:firstmomidealeqs}
\ea
The first equation above describes how the energy density and pressure evolve in response to fluid flow and the second equation describes how the fluid four-velocity responds to pressure gradients.

At this point, however, we have a small problem since we have more unknowns than equations.  The five unknowns are ${\cal E}$, ${\cal P}$, and the three independent components of $u^\mu$, but we only have four equations.  To close the system of equations, we must provide the relationship between ${\cal E}$ and ${\cal P}$ by imposing an equation of state (EoS).  This can be formulated as a constraint on the trace of the energy momentum tensor, ${T^\mu}_\mu = {\cal E} - 3 {\cal P} \equiv {\cal I}$, where ${\cal I}$ is called the trace-anomaly.  For a non-interacting ideal gas, ${\cal I}=0$ and we have ${\cal E} = 3 {\cal P}$.

Finally, let's consider a simple case of ideal hydrodynamical expansion that was originally presented by Bjorken \cite{Bjorken:1982qr} in order to get a feeling for how the temperature evolves in a heavy-ion collision.  The case we will consider is a boost-invariant system that is transversally homogenous (no transverse dynamics).  In this case, it is convenient to use comoving ``Milne'' coordinates 
\begin{eqnarray}
t&=&\tau \cosh\varsigma \, , \nonumber \\
z&=&\tau \sinh\varsigma \, . 
\label{eq:com-coord}
\end{eqnarray}
In this coordinate system, one as $\tilde{x}^\mu = (\tau,x,y,\varsigma)$ and the metric is ${\tilde g}_{\mu\nu}= {\rm diag}\,(1,-1,-1,-\tau^2)$. For a boost-invariant system, the four-velocity in Minkowski space is
\begin{equation}
u^\mu = (\cosh\varsigma,0,0,\sinh\varsigma) \, ,
\end{equation}
where here $\mu \in \{ t,x,y,z \}$.  Transforming this to Milne coordinates one finds 
\begin{equation}
\tilde{u}^\mu = (1,0,0,0) \, ,
\end{equation}
with, now, $\mu \in \{ \tau,x,y,\varsigma \}$.  With this we have
\begin{eqnarray}
D &=& \partial_\tau \, , \nonumber \\
\theta &=& \frac{1}{\tau} \, .
\label{eq:ddefs2}
\end{eqnarray}
By applying the last two expressions to the zeroth moment of the Boltzmann equation (\ref{eq:zeromom}) for an isotropic plasma we obtain
\begin{equation}
\partial_\tau n = - \frac{n}{\tau} \, ,
\end{equation}
which has a solution of the form
\begin{equation}
n(\tau) = n_0\,\frac{\tau_0}{\tau} \, .
\end{equation}

If now we apply again the expressions given in Eq.~(\ref{eq:ddefs2}) to the first moment of the Boltzmann equation given by Eqs.~(\ref{eq:firstmomidealeqs1}) and (\ref{eq:firstmomidealeqs}) one finds
\begin{equation}
\partial_\tau{\cal E} + \frac{{\cal E} +{\cal P}}{\tau} = 0 \, .
\end{equation}
If the system has an ideal EoS, then ${\cal E} = 3{\cal P}$, and one can further simplify this to
\begin{equation}
\partial_\tau{\cal E} = - \frac{4}{3} \frac{{\cal E}}{\tau} \, ,
\end{equation}
which has a solution
\begin{equation}
{\cal E}_{\rm ideal\;gas} = {\cal E}_0 \left(\frac{\tau_0}{\tau}\right)^{4/3} \, .
\end{equation}
If the system does not have an ideal EoS but instead has an equation of state corresponding to 
a constant speed of sound, i.e. $d{\cal P}/d{\cal E} = c_s^2$, then it follows that ${\cal P} = c_s^2 {\cal E}$
where we have fixed the constant by demanding that the pressure goes to zero when the energy density goes
to zero.  In this case one finds instead 
\begin{equation}
{\cal E} = {\cal E}_0 \left(\frac{\tau_0}{\tau}\right)^{1+c_s^2} \, ,
\end{equation}
which reduces to the ideal case when $c_s^2 = 1/3$.  If the EoS has  varying speed of sound then one
can express ${\cal P}$ in terms of an integral of the speed of sound. 

\section{Lecture 2}

In lecture 2 I will discuss how to include non-equilibrium corrections in the framework of hydrodynamics, first in the context of linearized second-order viscous hydrodynamics, and then in the context of leading order anisotropic hydrodynamics.

\subsection{Second-order viscous hydrodynamics}

The most commonly used method for including non-perturbative (and necessarily anisotropic) corrections to the ideal hydrodynamics equations obtained at the end of the last lecture is to expand the energy-momentum tensor as an ideal tensor, $T^{\mu\nu}_{\rm ideal}$ plus a tensor correction $\Pi^{\mu\nu}$.  It is typically implicitly assumed that all components of $\Pi^{\mu\nu}$ are small corrections to the leading-order ideal energy-momentum tensor
\be
T^{\mu\nu} = T^{\mu\nu}_{\rm ideal} + \Pi^{\mu\nu} \, .
\label{eq:Tmunusplit}
\ee
In general, one can decompose $\Pi^{\mu\nu}$ into a traceless part called $\pi^{\mu\nu}$ and a traceful part proportional to ${\rm tr}\,\Pi^{\mu\nu} = g_{\mu\nu} \Pi^{\mu\nu} = {\Pi^\mu}_\mu$
\be
\Pi^{\mu\nu} = \pi^{\mu\nu} - \Phi \Delta^{\mu\nu} \, .
\ee
where ${\pi^\mu}_\mu = 0$.  Using ${\Delta^\mu}_\mu = -3$, we obtain ${\Pi^\mu}_\mu = 3 \Phi$.  With this we can generalize Eq.~(\ref{eq:firstmomidealeqs}) to  
\ba
D {\cal E} + ({\cal E} + {\cal P}) \theta - \Pi^{\mu\nu} \nabla_{(\mu}u_{\nu)} &=& 0 \, ,
\\
({\cal E} + {\cal P}) D u^i - \nabla^i {\cal P} + {\Delta^i}_\nu \partial_\mu \Pi^{\mu\nu} &=& 0 \, ,
\label{eq:firstmomsecondordereqs}
\ea
where $A_{(\mu}B_{\nu)} \equiv (A_\mu B_\nu + A_\nu B_\mu)/2$ gives the symmetric part of a rank-two tensor.  In order to solve these equations, however, we need to know the full spatiotemporal evolution of $\Pi^{\mu\nu}$.  To do this in the kinetic theory framework, one must take projections of the second moment of the Boltzmann equation.

Before presenting the second-order result, we can consider the first-order result, which is given by
\ba
\pi^{\mu\nu}_{\rm NS} &=& 2 \eta \nabla^{\langle\mu} u^{\nu\rangle} \, ,
\\
\Phi_{\rm NS} &=& \zeta \nabla_\alpha u^\alpha \, ,
\ea
where $\nabla^{\langle\mu} u^{\nu\rangle} = \nabla^{(\mu} u^{\nu)} - \tfrac{1}{3} \Delta^{\mu\nu} \nabla_\alpha u^\alpha$ is the symmetric and traceless part of the fluid gradients, the coefficient $\eta$ is called the shear viscosity, and $\zeta$ is called the bulk viscosity.  Note that one can also introduce a four-index projector $\Delta^{\mu\nu}_{\alpha\beta}\equiv\Delta^{(\mu}_\alpha\Delta^{\nu)}_\beta-\Delta^{\mu\nu}\Delta_{\alpha\beta}/3$, which allows us to write $\nabla^{\langle\mu} u^{\nu\rangle} = \Delta^{\mu\nu}_{\alpha\beta} \nabla^\alpha u^\beta$ and, more generally, we can define $A^{\langle \mu \nu\rangle}\equiv\Delta^{\mu\nu}_{\alpha\beta}A^{\alpha\beta}$.

Before proceeding further, let's look at the number of degrees of freedom that are added when we include the non-equilibrium degrees of freedom.  In general, $T^{\mu\nu}$ is a symmetric tensor.  As a consequence, a bulk description of the energy-momentum tensor has 10 independent degrees of freedom.  As we have already discussed, the flow velocity $u^\mu$ has three independent components.  In order to make the separation between the ideal and viscous contributions precise, one typically requires that $u^\mu$ corresponds to the time-like eigenvector of the full energy-momentum tensor, i.e.
\be
u_\mu T^{\mu\nu} = {\cal E} u^\nu \, .
\ee
This choice is referred to as the Landau frame.  Since $u_\mu T^{\mu\nu}_{\rm ideal} = {\cal E} u^\nu$, this implies that $u_\mu \Pi^{\mu\nu} = 0$.  This implies that, in the local rest frame where $u^\mu=(1,0,0,0)$, $\Pi^{0\nu}_{\rm LRF} = \Pi^{\nu 0}_{\rm LRF} = 0$.  As result, there are only 6 possible independent components of $\Pi^{\mu\nu}$.  There will be 5 degrees of freedom contained in $\pi^{\mu\nu}$, since it is traceless, and 1 more coming from the traceful part which is proportional to $\Phi$.  Combining these with ${\cal E}$ and ${\cal P}$ and the three independent components of the fluid four-velocity, we obtain a grand total of 11 degrees of freedom.\footnote{For simplicity, in this analysis I have ignored the diffusion current and hence heat flow.  The equations of motion for this come from the zeroth moment of the Boltzmann equation.}  As before, one of these degrees of freedom is eliminated by imposing the relation between ${\cal E}$ and ${\cal P}$ implied by the EoS.  Therefore, we are left with 10 degrees of freedom as expected.  Since the ideal equations of motion provide 4 equations, we will need 6 more equations from the second-moment of the Boltzmann equation.

In order to determine these equations in the kinetic field theory framework, one uses the defining relation for the energy-momentum tensor in terms of the one-particle distribution function
\be
T^{\mu\nu}(x) = \int dP \, p^\mu p^\nu f(x,p) \, ,
\ee
where $dP = d^3p/(2\pi)^3 E$ is the Lorentz-invariant integration measure.  If we linearize the one-particle distribution function around an isotropic equilibrium distribution function using
\be
f(x,p) = f_{\rm eq}\left(\frac{p_\mu u^\mu}{T}\right) + \delta f(p,x) \, ,
\ee
we obtain
\be
T^{\mu\nu}(x) = T^{\mu\nu}_{\rm ideal}(x) + \int dP \, p^\mu p^\nu \delta f(x,p) \, .
\ee
Comparing this to Eq.~(\ref{eq:Tmunusplit}) allows us to identify
\be
\Pi^{\mu\nu} = \int dP \, p^\mu p^\nu \delta f(x,p) \, ,
\ee
Projecting out the symmetric and traceless part using $\Delta^{\mu\nu}_{\alpha\beta}$ we obtain
\be
\pi^{\mu\nu} = \int dP \, p^{\langle\mu} p^{\nu\rangle} \delta f(x,p) \, ,
\label{eq:pidfrel}
\ee
and taking the trace, we obtain
\be
\Phi = -\frac{1}{3} \int dP \, p^\mu p_\mu \delta f(x,p) \, .
\ee
From the second expression, we see that for a system of massless particles, for which $p^\mu p_\mu = 0$, one has $\Phi=0$.  In what follows in this lecture, I will restrict our considerations to the case of massless particles, but I note that recently there have been studies of the effect of bulk viscosity in rapidly expanding massive gases using the kinetic theory framework \cite{Denicol:2014vaa,Denicol:2014mca,Jaiswal:2014isa}

Specializing to the case that the equilibrium distribution function is a Boltzmann distribution, $f_{\rm eq}(x) = \exp(-x)$, we can invert Eq.~(\ref{eq:pidfrel}) to obtain $\delta f$ in terms of the shear tensor
\be
f(x,p) = f_{\rm eq}\!\left(\frac{p_\mu u^\mu}{T}\right) \left[1 + \frac{p^\alpha p^\beta \pi_{\alpha\beta}}{2({\cal E}+{\cal P})T^2} + {\cal O}\!\left(\frac{|p|^4|\pi_{\mu\nu}|^2}{T^4({\cal E}+{\cal P})^2} \right) \right] .
\ee
As one can see from this expression, there will be large corrections to the equilibrium distribution function in regions of phase space when $|p|/T > ({\cal E}+{\cal P})/|\pi_{\mu\nu}|$.  In order to get a feeling for where the troublesome regions in phase space are, we can consider the first-order approximation $\pi^{\mu\nu}_{\rm NS} = 2 \eta \nabla^{\langle\mu} u^{\nu\rangle}$ for the case of 0+1d expansion, in which case one finds that, in the local rest frame, $\pi^{xx}=\pi^{yy}=2\eta/3\tau$ and $\pi^{zz} =-4\eta/3\tau$.  In addition, if we work at zero chemical potential, we can use ${\cal E}+{\cal P}=T{\cal S}$ where ${\cal S}$ is the entropy density.  In this case, the expansion of the distribution function becomes
\be
f^{\rm 0+1d}_{\rm NS}(x,p) = f_{\rm eq}\!\left(\frac{E}{T}\right) \left[1 + \bar\eta \left( \frac{\hat{p}_x^2+\hat{p}_y^2-2\hat{p}_z^2}{3 \tau T} \right) + \cdots \right] ,
\label{eq:0p1nsf}
\ee
where $\bar\eta = \eta/{\cal S}$, $\hat{p} = p/T$, and the energy and all momenta are evaluated in the local rest frame.  From this expression, we learn that the overall magnitude of the correction is proportional to the ratio of the shear viscosity to entropy density, inversely proportional to $\tau T$, and anisotropic in momentum space, with the magnitude of the correction increasing quadratically in the magnitude of $\hat{p}$.  In fact, from this expression we see that there are regions of phase space where $f^{\rm 0+1d}_{\rm NS}(x,p) < 0$.

\begin{figure}[t]
\centerline{\includegraphics[width=0.6\linewidth]{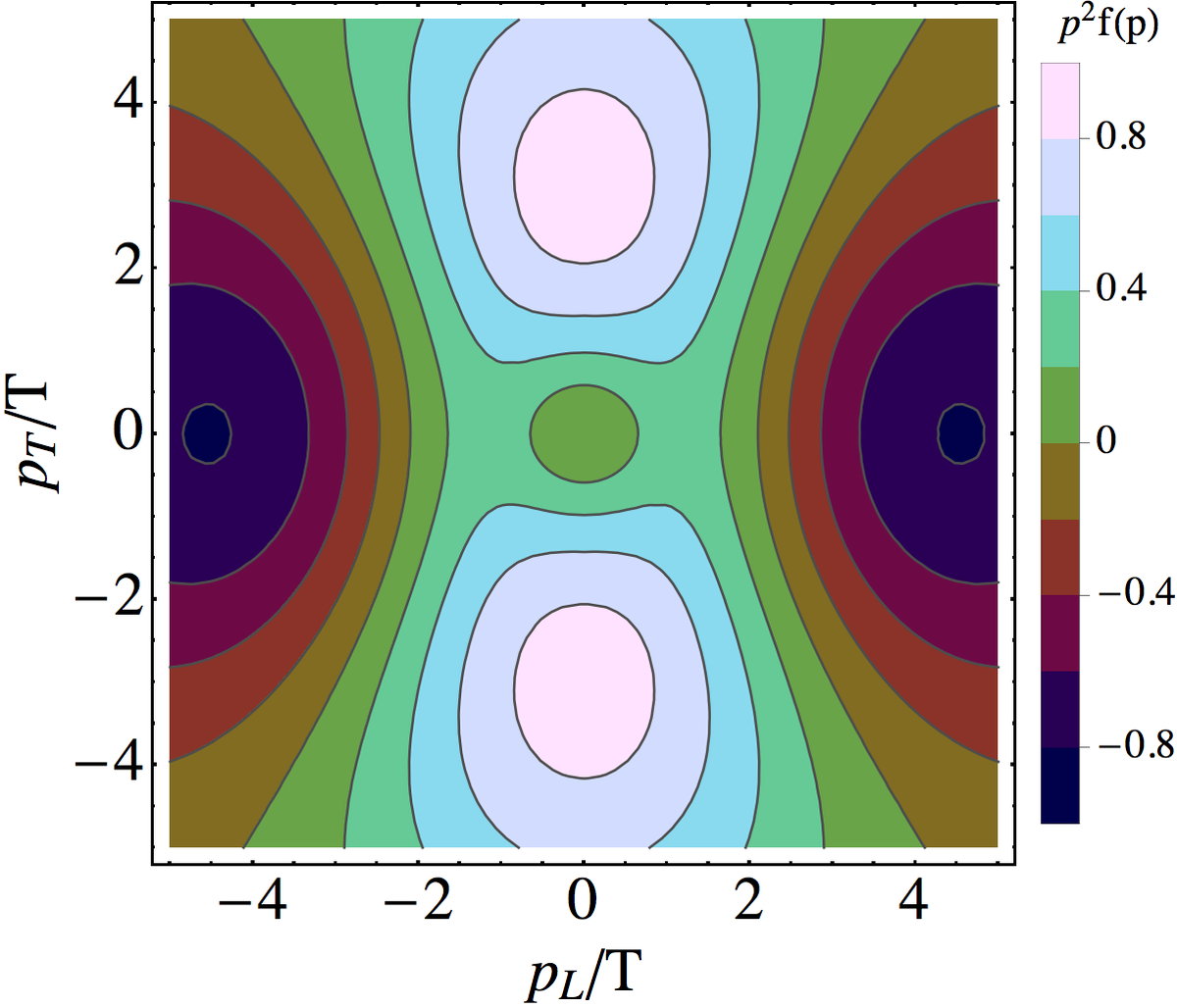}}
\caption{(Color online) Contour plot of $p^2 f^{\rm 0+1d}_{\rm NS}$ given by Eq.~(\ref{eq:0p1nsf}) using $4\pi\bar\eta = 3$ and $T = 230$ MeV at $\tau=0.6$ fm/c.}
\label{fig:contour}
\end{figure}

To quantify this further, let's take $4\pi\bar\eta = 3$ and $T = 230$ MeV, which corresponds to the conditions generated in the lower right panel of Fig.~\ref{fig:nscomp} at $\tau=0.6$ fm/c.  In Fig.~\ref{fig:contour}, I present a contour plot of $p^2 f^{\rm 0+1d}_{\rm NS}$. The factor of $p^2$ takes into account the phase factor that appears in the integral that defines the number density in spherical momentum-coordinates.  As we can see from this figure, there are large regions where the distribution function is negative.  If one were to use such a linearly-corrected distribution function for the calculation of, e.g., photon production, this would result in a relatively large error in the calculation, since the distribution function should never be negative.  It would be nice to have a formalism that, at leading order guarantees that the distribution is always $\geq 0$ such as anisotropic hydrodynamics.  Before proceeding to this, however, let's return to the development of relativistic viscous hydrodynamics.

As I said previously, the introduction of non-equilibrium corrections to ideal hydrodynamics requires additional equations of motion.  At first order the Naiver-Stokes solution is acausal and one has to go to second order in gradients in order have a causal theory of relativistic hydrodynamics.  The equation of motion for the shear tensor obtained from the second moment of the Boltzmann equation using a variant of the M\"uller-Israel-Stewart second-order formalism is~\cite{Romatschke:2009im}
\be
\pi^{\mu\nu} + \tau_\pi \! \left[ {\Delta^\mu}_\alpha {\Delta^\nu}_\beta \pi^{\alpha\beta} + \frac{4}{3} \pi^{\mu\nu} \nabla_\alpha u^\alpha - 2 \pi^{\alpha(\mu}{\Omega^{\nu)}}_{\!\alpha}
+ \frac{\pi^{\alpha\langle\mu}{\pi^{\nu\rangle}}_{\!\alpha}}{\eta} \right]\! = 2 \eta \nabla^{\langle\mu} u^{\nu\rangle} ,
\ee
where $\Omega_{\alpha\beta} = \tfrac{1}{2} (\nabla_\alpha u_\beta - \nabla_\beta u_\alpha)$ and $\tau_\pi$ is the shear relaxation time.  The most important feature of the above equation is the appearance of the shear relaxation time, $\tau_\pi$.  If we take $\tau_\pi=0$, we recover the first-order Navier-Stokes result, however, for any finite $\tau_\pi$, the theory will be casual.  The shear relaxation time, $\tau_\pi$, sets the timescale for the second-order solution for the shear tensor to approach the Navier-Stokes solution.  Since the Navier-Stokes solution is inherently momentum-space anisotropic, one can interpret $\tau_\pi$ as the ``anisotropization'' time scale.  In the strong-coupling limit of ${\cal N}=4$ SYM one finds $\tau_\pi = (2 - \log 2)/2 \pi T$ \cite{Baier:2007ix,Bhattacharyya:2008jc} which gives $\tau_\pi \sim 0.1$ fm/c and $\tau_\pi \sim 0.07$ fm/c for the RHIC- and LHC-like initial conditions stated in Lecture 1, respectively.  Therefore, one expects to see very rapid anisotropization of the QGP generated in a heavy-ion collision.

\subsection{Leading-order anisotropic hydrodynamics}

Since one expects to see rapid anisotropization of the QGP generated in heavy-ion collisions, it might be efficacious to take into account the existence of these  momentum-space anisotropies from the outset.  As discussed in the introduction, this can be accomplished by generalizing the leading-order term in the expansion of the one-particle distribution function to
\begin{equation}
   f(x,p) = 
   f_{\rm iso}\!\left(\frac{\sqrt{p^\mu \Xi_{\mu\nu}(x)p^\nu}}{\Lambda(x)}, \frac{\mu(x)}{\Lambda(x)}\right) \, .
\label{eq:ahexp2}
\end{equation}
\begin{figure}[t]
\centerline{\includegraphics[width=0.75\linewidth]{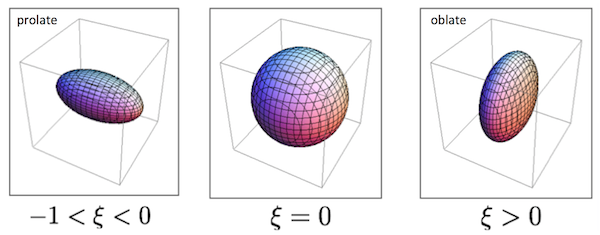}}
\caption{(Color online) Equal occupation number surface for the spheroidal anisotropic hydrodynamics distribution function.}
\label{fig:spheroids}
\end{figure}
The original formulation of anisotropic hydrodynamics was based on an azimuthally symmetric (spheroidal) ansatz for the local rest frame (LRF) anisotropy tensor $\Xi_{\mu\nu}(x)$ \cite{Martinez:2010sc,Florkowski:2010cf}.  In this case, the anisotropy tensor only involves a single anisotropy parameter $\xi$ with $\Xi^{\mu\nu}_{\rm LRF}(x) = {\rm diag}(1,0,0,\xi(x))$ and, therefore, for a system of massless particles, $p^\mu \Xi_{\mu\nu}(x)p^\nu$ reduces to ${\bf p}^2 + \xi(x) p_L^2$ in the LRF.  In the spheroidal formulation, $\xi=0$ gives an isotropic distribution, $-1 < \xi < 0$ gives a prolate distribution, and $0 < \xi < \infty$ gives an oblate distribution (see Fig.~\ref{fig:spheroids}).  We will take this as the definition of leading-order (LO) anisotropic hydrodynamics for the remainder of this lecture.  In the next lecture, we will discuss possible generalizations of the leading-order anisotropy tensor.

\begin{figure}[t]
\centerline{\includegraphics[width=0.8\linewidth]{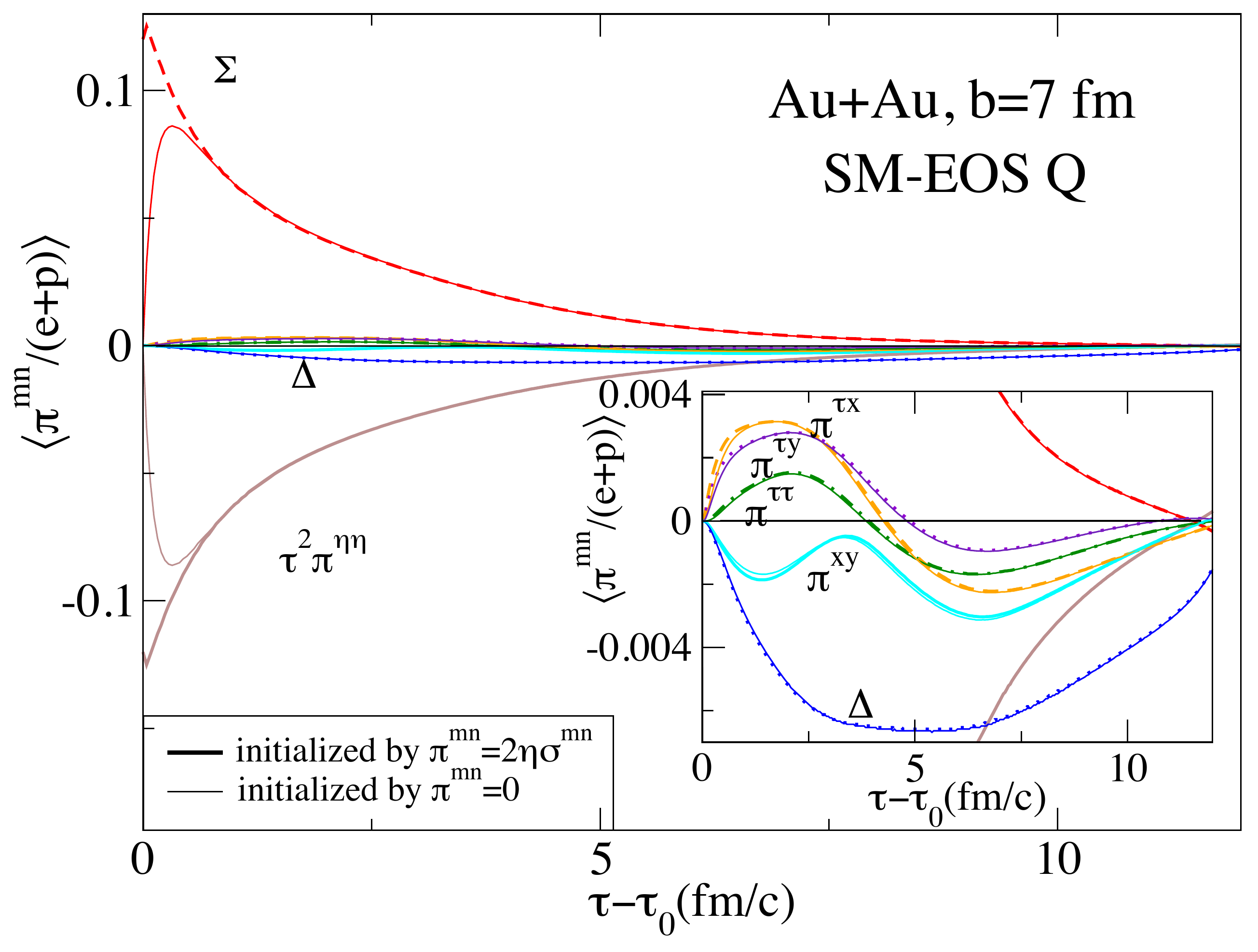}}
\caption{(Color online) Proper-time evolution of the components of the shear tensor obtained from a realistic second-order viscous hydrodynamics simulation with impact parameter $b=7$.  Figure taken from the PhD dissertation of H. Song \cite{Song:2009gc}.}
\label{fig:songdiss}
\end{figure}

In order to motivate why a spheroidal form might be a good starting point, in Fig.~\ref{fig:songdiss}, I present a plot made by Huichao Song in her PhD dissertation.  The figure shows the proper-time evolution of all of the components of the shear tensor obtained from a realistic second-order viscous hydrodynamics simulation.  As can be seen from this figure, two of the components plotted are much larger than the rest.  These correspond to the sum of the spacelike components $\Sigma \equiv \pi^{xx} + \pi^{yy}$ and $\tau^2 \pi^{\eta\eta} = \pi^{zz}$.\footnote{Here $\eta$ is the spatial rapidity.}  The quantity $\Sigma/2$ gives the viscous correction to the transverse pressure and $\pi^{zz}$ gives the viscous correction to the longitudinal pressure.  The next smallest thing plotted in Fig.~\ref{fig:songdiss} is the difference $\Delta \equiv \pi^{xx} - \pi^{yy}$, which is smaller than $\Sigma$ and $\pi^{zz}$ up to times on the order of 7 fm/c.  This means that, to very good approximation, one can treat the difference between $\pi^{xx}$ and $\pi^{yy}$ as a perturbation.  Likewise, we see that all off-diagonal components are even smaller.  So small, in fact, that they require a zoomed inset to visualize.  Once again this suggests that one can treat these components perturbatively.  At leading-order, therefore, a good approximation might be to assume that the distribution function, and hence the shear corrections, are spheroidal in form and treat the evolution of these, potentially large, corrections non-perturbatively.

Another benefit of the spheroidal form is that, for a massless gas, one can evaluate all components of the energy-momentum tensor analytically, with the non-vanishing components in Milne coordinates being~\cite{Martinez:2009ry,Martinez:2010sc}
\ba
{\cal E}(\Lambda,\xi) &=& T^{\tau\tau} = {\cal R}(\xi)\,{\cal E}_{\rm iso}(\Lambda) \, , 
\label{eq:ahed}
\\
{\cal P}_T(\Lambda,\xi) &=& \frac{1}{2}\left( T^{xx} + T^{yy}\right) = {\cal R}_{\rm T}(\xi){\cal P}_{\rm iso}(\Lambda) \, ,
\label{eq:ahpt}
\\
{\cal P}_L(\Lambda,\xi) &=& - T^{\varsigma}_\varsigma = {\cal R}_{\rm L}(\xi){\cal P}_{\rm iso}(\Lambda) \, ,
\label{eq:ahpl}
\ea
where ${\cal E}_{\rm iso}(\Lambda)$ and ${\cal P}_{\rm iso}(\Lambda)$ are the isotropic energy density and pressure computed in the isotropic limit using $f_{\rm iso}$ and
\ba
{\cal R}(\xi) &=& \frac{1}{2}\left(\frac{1}{1+\xi}
+\frac{\arctan\sqrt{\xi}}{\sqrt{\xi}} \right) ,
\label{eq:rfunc}
\\
{\cal R}_{\rm T}(\xi) &=& \frac{3}{2 \xi} 
\left( \frac{1+(\xi^2-1){\cal R}(\xi)}{\xi + 1}\right) ,
\\
{\cal R}_{\rm L}(\xi) &=& \frac{3}{\xi} 
\left( \frac{(\xi+1){\cal R}(\xi)-1}{\xi+1}\right) ,
\ea
which satisfy ${\cal R} = 2 {\cal R}_T + {\cal R}_L$.  As the expressions above show, for a  massless gas, the spheroidal energy density, transverse pressure, and longitudinal pressure all factorize multiplicatively into a function that only depends on the anisotropy and a function that only depends on the momentum scale.  This is important, because it allows us to impose the EoS as a relationship between ${\cal E}_{\rm iso}(\Lambda)$ and ${\cal P}_{\rm iso}(\Lambda)$ and then the extension to an "anisotropic EoS" is automatically taken care of by the ${\cal R}$, ${\cal R}_T$, and ${\cal R}_L$ functions.  I also note for completeness that a similar factorization occurs for the spheroidal number density
\be
n(\Lambda,\xi) = \frac{n_{\rm iso}(\Lambda)}{\sqrt{1+\xi}} \, .
\label{eq:ahn}
\ee
In addition, for a massless gas, one finds that using a spheroidal distribution function, all higher-order moments can also be computed analytically \cite{Bazow:2013ifa}.

\mysubsubsection{General tensor basis}

To proceed systematically, we should go back to the beginning and establish a tensor basis that can be used in general and then restrict to a spheroidal form.  A completely general tensor basis can be constructed by introducing four 4-vectors which in the LRF are \cite{Florkowski:2011jg,Martinez:2012tu}
\begin{eqnarray}
&&X^\mu_{0,{\rm LRF}} \equiv u^\mu_{\rm LRF} = (1,0,0,0) \nonumber \\
&&X^\mu_{1,{\rm LRF}} \equiv x^\mu_{\rm LRF} = (0,1,0,0) \nonumber \\
&&X^\mu_{2,{\rm LRF}} \equiv y^\mu_{\rm LRF} = (0,0,1,0) \nonumber \\
&&X^\mu_{3,{\rm LRF}} \equiv z^\mu_{\rm LRF} = (0,0,0,1) \, .
\label{eq:rfbasis}
\end{eqnarray}
These 4-vectors are orthonormal in all frames.  The vector $X^\mu_0$ is associated with the four-velocity of the local rest frame and is canonically called $u^\mu$ and one can also identify $X^\mu_1 = x^\mu$,  $X^\mu_2 = y^\mu$, and $X^\mu_3 = z^\mu$ as indicated above.  I will use the two different labels for these vectors interchangeably depending on convenience since the notation with numerical indices allows for more compact expressions in many cases.  Note that, in the lab frame the three spacelike vectors $X^\mu_i$ can be written entirely in terms of $X^\mu_0=u^\mu$.  This is because $X^\mu_i$ can be obtained by a sequence of Lorentz transformations/rotations applied to the local rest frame expressions specified above.

Finally, I point out that one can express the metric tensor itself in terms of these 4-vectors as
\begin{equation}
g^{\mu \nu}= X^\mu_0 X^\nu_0 - \sum_{i=1}^3 X^\mu_i X^\nu_i \, .
\label{eq:gbasis}
\end{equation}
In addition, the transverse projection operator, which is orthogonal to $X^\mu_0$, can be rewritten in terms of the vector basis (\ref{eq:rfbasis}) as
\begin{equation}
\Delta^{\mu \nu} = g^{\mu\nu} - X^\mu_0 X^\nu_0 = - \sum_{i=1}^3 X^\mu_i X^\nu_i \, ,
\label{eq:transproj}
\end{equation}
such that $u_\mu \Delta^{\mu \nu} = u_\nu \Delta^{\mu \nu} = 0$.  We note that the spacelike components of the tensor basis are eigenfunctions of this operator, i.e. $X_{i\mu} \Delta^{\mu \nu} = X^\nu_{i}$.

\mysubsubsection{Spheroidal anisotropic energy-momentum tensor}

Since the energy-momentum tensor is a symmetric rank-two tensor, we can express it generally as 
\begin{equation}
T^{\mu\nu}(t,{\bf x}) = t_{00} g^{\mu \nu}  + \sum_{i=1}^3 t_{ii} X^\mu_i X^\nu_i 
+ \sum_{\alpha,\beta=0 \atop \alpha>\beta}^3 t_{\alpha\beta} (X^\mu_\alpha X^\nu_\beta+X^\mu_\beta X^\nu_\alpha) \, ,
\label{eq:emtensorgen}
\end{equation}
In the case of spheroidal anisotropic hydrodynamics one has
\begin{eqnarray}
T^{00}_{\rm LRF} &=& {\cal E} = t_{00} \nonumber \, , \\
T^{xx}_{\rm LRF} &=& {\cal P}_\perp = -t_{00} + t_{11}\nonumber \, ,  \\
T^{yy}_{\rm LRF} &=& {\cal P}_\perp = -t_{00} + t_{22}\nonumber \, ,  \\
T^{zz}_{\rm LRF} &=&  {\cal P}_L = -t_{00} + t_{33} \, ,
\end{eqnarray}
and, due to the spheroidal symmetry in momentum-space, we must have $t_{11}=t_{22}$ which gives four equations for our four unknowns.  Solving for the coefficients $t_{\mu\nu}$ and relabeling $X^\mu_0 \rightarrow u^\mu$ and $X^\mu_3 \rightarrow z^\mu$, we obtain
\begin{equation}
T^{\mu\nu} = ({\cal E} +{\cal P}_\perp) u^\mu u^\nu  - {\cal P}_\perp g^{\mu \nu}+ ({\cal P}_L - {\cal P}_\perp) z^\mu z^\nu \, ,
\label{eq:speroidaltmunu}
\end{equation}
which, in the isotropic limit with ${\cal P}_\perp = {\cal P}_L \equiv {\cal P}$, reduces to (\ref{eq:tmunuideal2}).

\mysubsubsection{Equations of motion}

Using the above tensor basis and taking the zeroth and first moments of the Boltzmann equation, one can derive the equations of motion.  If one uses a relaxation time approximation collisional kernel, the 2+1d aniostropic hydrodynamics equations of motion appropriate for describing the spatiotemporal evolution of a boost-invariant system are~\cite{Martinez:2012tu}
\begin{equation}
\frac{1}{1+\xi}D\xi - 6D(\log\Lambda) - 2 \theta = 2 \Gamma\left(1 - {\cal R}^{3/4}(\xi) \sqrt{1+\xi}\right) ,
\label{eq:zeromomeq}
\end{equation}
and
\begin{eqnarray}
{\cal R}'(\xi) D\xi + 4 {\cal R}(\xi) D(\log\Lambda) &=& 
- \left({\cal R}(\xi) + \frac{1}{3} {\cal R}_\perp(\xi)\right) \Delta_\perp
\nonumber \\
&& \hspace{1cm} - \left({\cal R}(\xi) + \frac{1}{3} {\cal R}_L(\xi)\right) \frac{u_0}{\tau} \, ,
\\
\left[3{\cal R}(\xi) + {\cal R}_\perp(\xi)\right] D u_\perp &=&-u_\perp \left[ {\cal R}_\perp'(\xi) \tilde{D} \xi 
+ 4  {\cal R}_\perp(\xi) \tilde{D} (\log\Lambda) \right.
\nonumber \\
&& \hspace{1cm} \left. + \frac{u_0}{\tau} ({\cal R}_\perp(\xi)-{\cal R}_L(\xi)) \right] ,
\\
u_y^2 \left[3{\cal R}(\xi) + {\cal R}_\perp(\xi)\right] D \left( \frac{u_x}{u_y} \right) &=& 
{\cal R}_\perp'(\xi) D_\perp\xi 
+ 4 {\cal R}_\perp(\xi) D_\perp(\log\Lambda) \, ,
\label{eq:firstmomeqs}
\end{eqnarray}
where
\begin{eqnarray}
\Delta_\perp &\equiv& \partial_\tau u_0 + \nabla_\perp \cdot {\bf u}_\perp \, , \nonumber \\
\tilde{D} &\equiv& u_0 \partial_\tau + \frac{u_0^2}{u_\perp^2} {\bf u}_\perp \cdot \nabla_\perp \, , \nonumber \\
D_\perp &\equiv& \hat{\bf z} \cdot ({\bf u}_\perp \times \nabla_T) = u_x \partial_y - u_y \partial_x \, ,
\end{eqnarray}
with ${\bf u}_\perp \equiv (u_x,u_y)$, and $u_0^2 = 1 + u_\perp^2$.

\begin{figure}
\begin{center}
\hspace{-1mm}\includegraphics[width=0.35\linewidth]{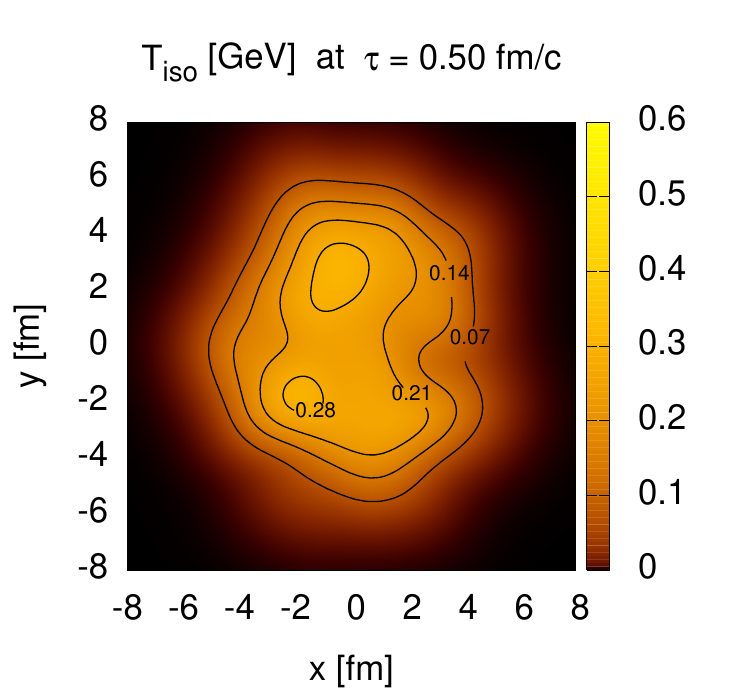}\hspace{-4mm}
\includegraphics[width=0.35\linewidth]{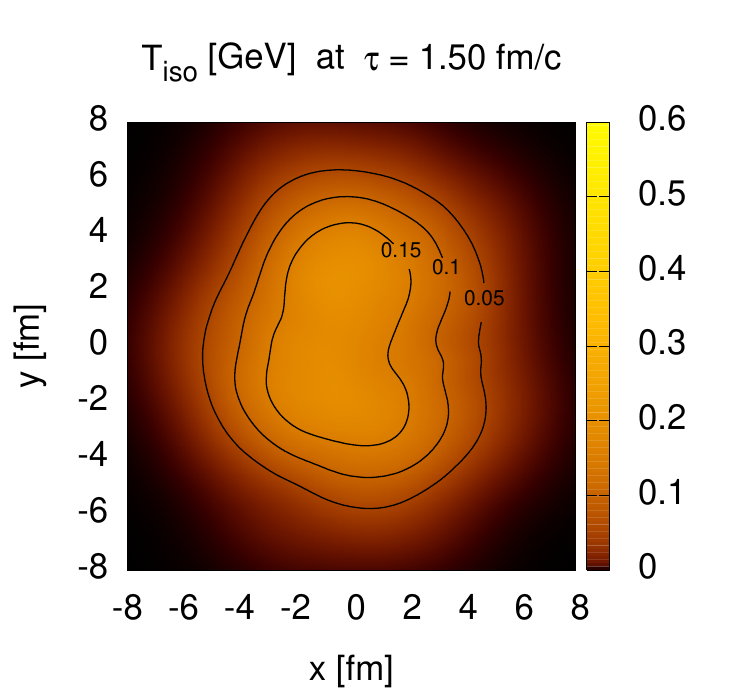}\hspace{-4mm}
\includegraphics[width=0.35\linewidth]{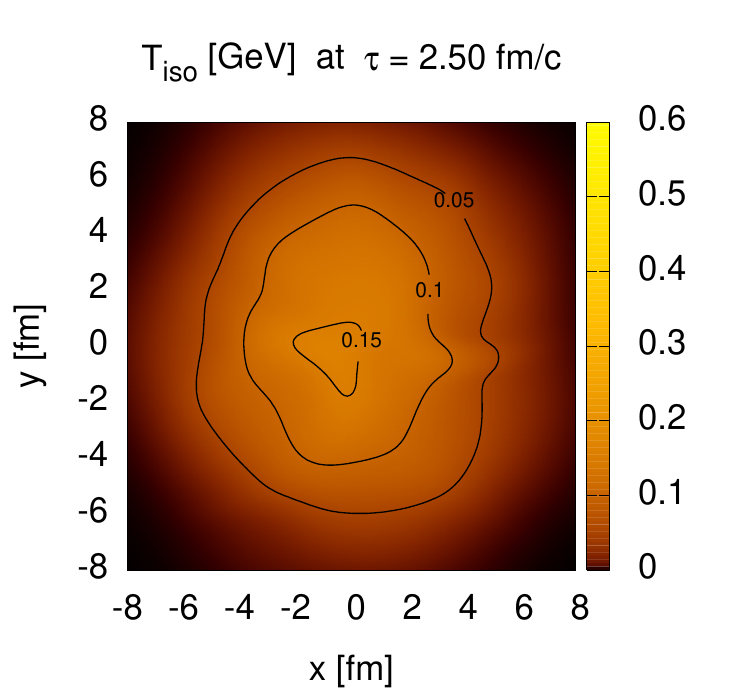}\\
\hspace{-1mm}\includegraphics[width=0.35\linewidth]{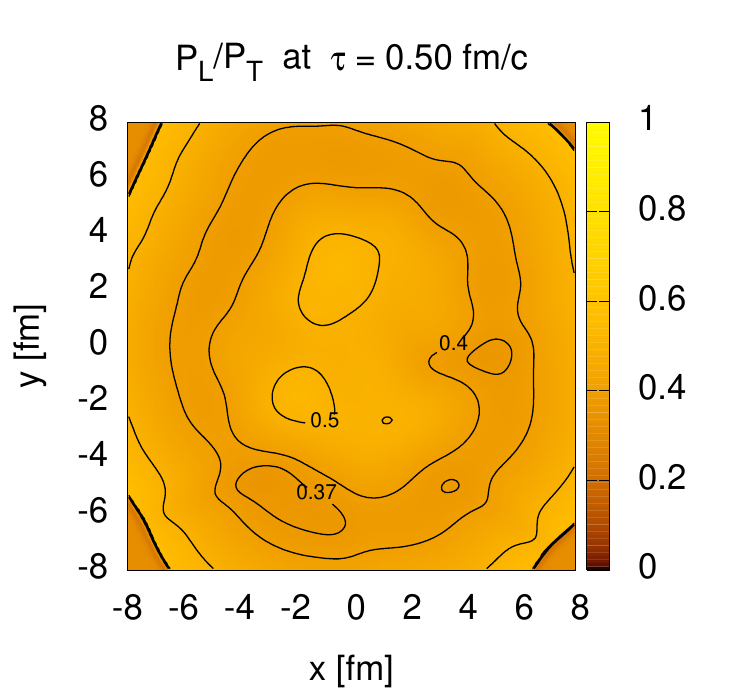}\hspace{-4mm}
\includegraphics[width=0.35\linewidth]{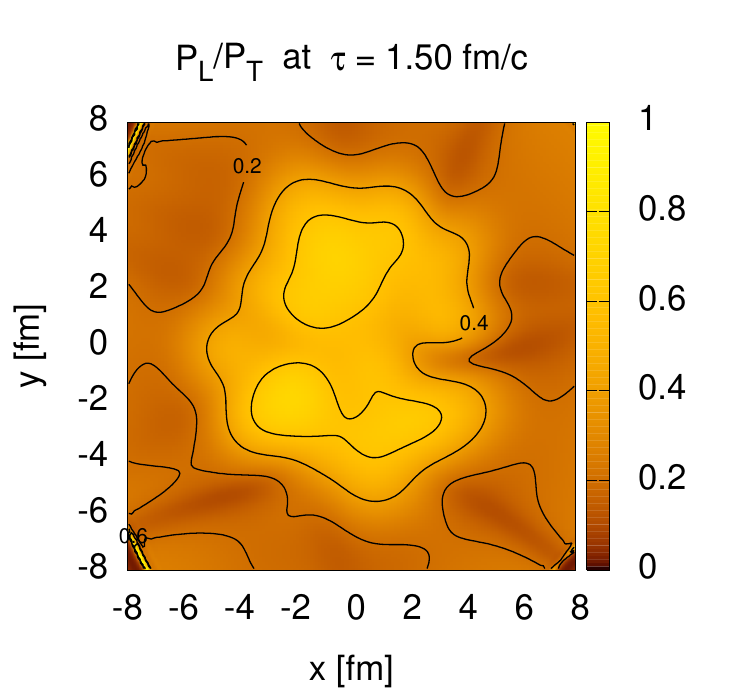}\hspace{-4mm}
\includegraphics[width=0.35\linewidth]{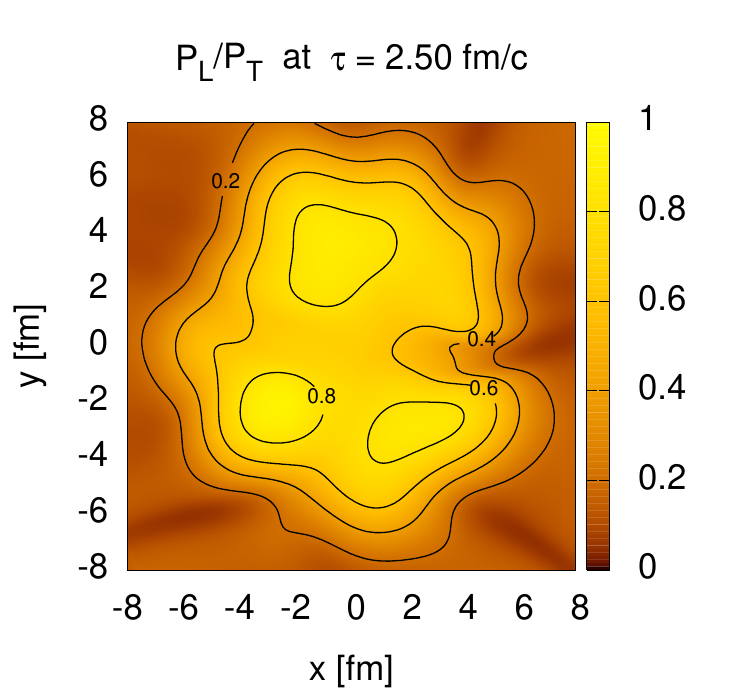}
\end{center}
\caption{(Color online) Visualization of the effective temperature and pressure anisotropy at three different times after the
nuclear impact.  For these plots we assumed a collision centrality of $b=7$ fm with a sampled Monte-Carlo Glauber wounded-nucleon profile.  The initial isotropic temperature for a central collision was taken to be $T=0.6$ GeV at $
\tau = 0.25$ fm/c. For this plot we used a value of $4\pi\eta/{\cal S} = 1$.  Figure taken from Ref.~\cite{Martinez:2012tu}.
}
\label{fig:b7MCvis}
\end{figure}

Equations~(\ref{eq:zeromomeq}) and (\ref{eq:firstmomeqs}) were numerically solved in Ref.~\cite{Martinez:2012tu}.  In Fig.~\ref{fig:b7MCvis}, I show a visualization of the dynamics for the case $4\pi\eta/{\cal S} = 1$.  The top row of  Fig.~\ref{fig:b7MCvis} shows the effective temperature obtained from the (scaled) fourth root of the energy density at three different proper times.  The bottom row of  Fig.~\ref{fig:b7MCvis} shows the ratio of the longitudinal pressure and transverse pressure at the same times.  As can be seen from this figure, even in the case that $4\pi\eta/{\cal S} = 1$, a high degree of momentum-space anisotropy is generated.  In addition, we see that, for the case of fluctuating initial conditions, there can be regions with a high degree of momentum-space anisotropy even in the center of the simulation.  Finally, I note that one sees that the pressures (and also one-particle distribution functions) are positive everywhere, even in the extremely dilute/low temperature region, where there are large non-equilibrium corrections.

\section{Lecture 3}

In the previous lecture I reviewed the derivation of second-order viscous hydrodynamics which motivated the development of the anisotropic hydrodynamics framework.  In this lecture I would like to begin by addressing the question of how one can determine if one hydrodynamical framework is better than another.  For this purpose I will discuss recently obtained exact solutions of the relaxation time approximation (RTA) Boltzmann equation.  Finally, I will give a brief review of what has been accomplished in the context of anisotropic hydrodynamics in the last year.

\subsection{Exact solutions to the RTA Boltzmann equation}

In order to judge the efficacy of different hydrodynamics frameworks, it would be nice to have some exactly solvable cases with which to compare the various approximations.  With this in mind, recently Florkowski et al. have exactly solved the Boltzmann equation for a transversely homogeneous boost-invariant system in the relaxation time approximation (RTA) \cite{Florkowski:2013lza,Florkowski:2013lya,Florkowski:2014sfa}.  

The starting point for the solution is the RTA Boltzmann equation $p^\mu \partial_\mu f(x,p) = -C[f(x,p)]$ with 
\begin{equation}
C[f] = \frac{p_\mu u^\mu}{\tau_{\rm eq}} \biggl[ f(x,p) - f_{\rm eq}\Big(p_\mu u^\mu,T(x)\Big) \biggr],
\label{eq:col-term}
\end{equation}
where $u^\mu$ is the local rest frame four velocity, $\tau_{\rm eq}$ is the relaxation time which may depend on proper time, and $f_{\rm eq}$ is an equilibrium distribution function that may be taken to be a Bose-Einstein, Fermi-Dirac, or Boltzmann distribution.  The effective temperature $T(\tau)$ appearing in the argument of the equilibrium distribution function is fixed by requiring energy-momentum conservation \cite{Baym:1984np}.  In Refs.~\cite{Florkowski:2013lza,Florkowski:2013lya} Florkowski et al. restricted themselves to the case of massless particles and in Ref.~\cite{Florkowski:2014sfa} they extended the solution to the case of massive particles.  In these lectures, I will present the details only for the case of massless particles and refer the reader to Ref.~\cite{Florkowski:2014sfa} for the massive case.

In order to simplify the Boltzmann equation for a transversely homogeneous boost-invariant system, one can define variables $w =  t p_z - z E$ and $v = Et-p_z z$ \cite{Bialas:1984wv,Bialas:1987en,Florkowski:2012ax}.  When written in terms of these variables, the left hand side of the Boltzmann equation becomes simply $p^\mu \partial_\mu f = (v/\tau) \partial_\tau f$.  This allows one to solve the RTA Boltzmann equation exactly
\begin{equation}
f(\tau,w,p_\perp) = D(\tau,\tau_0) f_0(w,p_\perp)  
+ \int_{\tau_0}^\tau \frac{d\tau^\prime}{\tau_{\rm eq}(\tau^\prime)} \, D(\tau,\tau^\prime) \, 
f_{\rm eq}(\tau^\prime,w,p_\perp) \, ,  
\label{eq:solf}
\end{equation}
where $\tau_0$ is the initial proper time, $f_0$ is the initial non-equilibrium distribution function, and ${D(\tau_2,\tau_1) = \exp\!\left[-\int_{\tau_1}^{\tau_2} d\tau^{\prime\prime} \, \tau^{-1}_{\rm eq}(\tau^{\prime\prime})\right]}$ is the damping function.  This solution is similar to the one obtained originally by Baym \cite{Baym:1984np}, but has been extended to an arbitrary initial condition at $\tau_0\neq0$ and allows for the possibility that the relaxation time $\tau_{\rm eq}$ is time dependent.  In the relaxation time approximation, one finds $\tau_{\rm eq} = 5 \eta/(T {\cal S})$ where $\eta$ is the shear viscosity, ${\cal S}$ is the entropy density, and $T$ is the effective temperature which we will specify below \cite{Anderson1974466,Czyz:1986mr}.\footnote{I note that, when employing the Grad-Israel-Stewart approximation truncated at second order in moments, one finds instead $\tau_{\rm eq} = 6 \eta/(T {\cal S})$.  This is an artifact of an incomplete second order truncation.  The correct result is $\tau_{\rm eq} = 5 \eta/(T {\cal S})$.}  In Refs.~\cite{Florkowski:2013lza,Florkowski:2013lya}, Florkowski et al. assumed  that $\eta/ {\cal S}$ was time independent, however, the exact solution also allows for a temperature-dependent $\eta/{\cal S}$.

Based on Eq.~(\ref{eq:solf}), one can evaluate the energy density via
\begin{equation}
{\cal E}(\tau) = g \int dP \, v^2\,  f(\tau,w,p_\perp)/\tau^2 \, ,
\end{equation}
where $g$ is the degeneracy factor and $dP = 2 \, d^4p \, \delta(p^2) \theta(p^0) = v^{-1} \, dw \, d^2p_T$. Integrating Eq.~(\ref{eq:solf}), one obtains an integral equation for the energy density
\begin{eqnarray}
\bar{\cal E}(\tau) &=& D(\tau,\tau_0) \,
{\cal R}\big(\xi_{\rm FS}(\tau)\big)\big/{\cal R}\left(\xi_0\right)
\nonumber \\
&& \hspace{1cm}
+ \int_{\tau_0}^{\tau} \! \frac{d\tau^\prime}{\tau_{\rm eq}(\tau^\prime)} \, D(\tau,\tau^\prime)  \, {\cal R}\!\left( \! \left(\frac{\tau}{\tau^\prime}\right)^2 - 1 \right) 
\bar{\cal E}(\tau^\prime) ,
\label{eq:inteq}
\end{eqnarray}
where ${\bar{\cal E} = {\cal E}/{\cal E}_0}$ is the energy density scaled by the initial energy density, $\xi_0$ is the initial momentum-space anisotropy, ${\xi_{\rm FS}(\tau) = (1+\xi_0)(\tau/\tau_0)^2-1}$, and ${\cal R}$ was defined previously in Eq.~(\ref{eq:rfunc}).

Equation~(\ref{eq:inteq}) can be solved numerically using the method of iteration.  
From the resulting energy density, one can obtain the effective temperature via ${\cal E}(\tau) = \gamma \, T^4(\tau)$ where $\gamma$ is a constant which depends on the particular equilibrium distribution function assumed and the number of degrees of freedom.  
The resulting effective temperature allows one to determine the distribution function $f_{\rm eq}$ at all proper times and, with this, the full particle distribution function can be obtained using Eq.~(\ref{eq:solf}).
Additionally, one can determine the number density, longitudinal pressure, and transverse pressure by integrating the distribution function multiplied by $v/\tau$, $w^2/\tau^2$, and $p_T^2/2$, respectively \cite{Florkowski:2013lya}.

Florkowski et al. compared the exact solution with the LO spheroidal anisotropic hydrodynamics (AH) equations obtained from the zeroth and first moments of Boltzmann equation in RTA~\cite{Martinez:2010sc}
\begin{eqnarray}
\frac{1}{1+\xi} \partial_\tau \xi- \frac{2}{\tau} - \frac{6}{\Lambda} \partial_\tau \Lambda &=& 
\frac{2}{\tau_{\rm eq}^{\rm AH}} \left[ 1 - {\cal R}^{3/4}(\xi) \sqrt{1+\xi} \right] \! ,
\nonumber \\
\frac{{\cal R}'(\xi)}{{\cal R}(\xi)} \partial_\tau \xi + \frac{4}{\Lambda} \partial_\tau \Lambda &=& 
\frac{1}{\tau} \left[ \frac{1}{\xi(1+\xi){\cal R}(\xi)} - \frac{1}{\xi} - 1 \right] \! , \;\;\;\;
\label{eq:ahydro}
\end{eqnarray}
where $\Lambda$ is the transverse temperature and ${\tau_{\rm eq}^{\rm AH} = 5 \eta/(2 \Lambda {\cal S})}$ is the relaxation time.  The time evolution of $\xi$ and $\Lambda$ is obtained by solving Eqs.~(\ref{eq:ahydro}) and, using these, one can straightforwardly compute the time dependence of the energy density, transverse pressure, longitudinal pressure, and number density using Eq.~(\ref{eq:ahed}), (\ref{eq:ahpt}), (\ref{eq:ahpl}), and (\ref{eq:ahn})~\cite{Martinez:2009ry,Martinez:2010sc}.

In addition, they compared the exact solution with two second order viscous hydro prescriptions, both of which can be written compactly as 
\begin{eqnarray}
\partial_\tau {\cal E}&=&-\frac{{\cal E}+{\cal P}}{\tau}+\frac{\Pi}{\tau} \; ,
\nonumber \\
\partial_\tau\Pi&=&-\frac{\Pi}{\tau_\pi}+\frac{4}{3}\frac{\eta}{\tau_\pi\tau}-\beta\frac{\Pi}{\tau}\,,
\label{eq:vhydro}
\end{eqnarray}
where $\Pi = {\Pi^\varsigma}_\varsigma$ is the shear and $\tau_\pi = 5 \eta/(T {\cal S})$ is the shear relaxation time. In the majority of the literature, practitioners use $\beta = 4/3$ which we will refer to as the Israel-Stewart (IS) prescription.  
They also compared the exact solutions with the complete second order treatment from Ref.~\cite{Denicol:2012cn} which, within the relaxation time approximation, gives $\beta = 38/21$.  We will refer to the second choice as the DNMR prescription.\footnote{Reference~\cite{Jaiswal:2013npa} has also obtained $\lambda = 38/21$ using the Chapman-Enskog method.}  In both cases one can compute the transverse pressure via ${\cal P}_T = {\cal P} + \Pi/2$ and the longitudinal pressure via ${\cal P}_L = {\cal P} - \Pi$.

\begin{figure}[t]
\begin{center}
\includegraphics[width=0.48\linewidth]{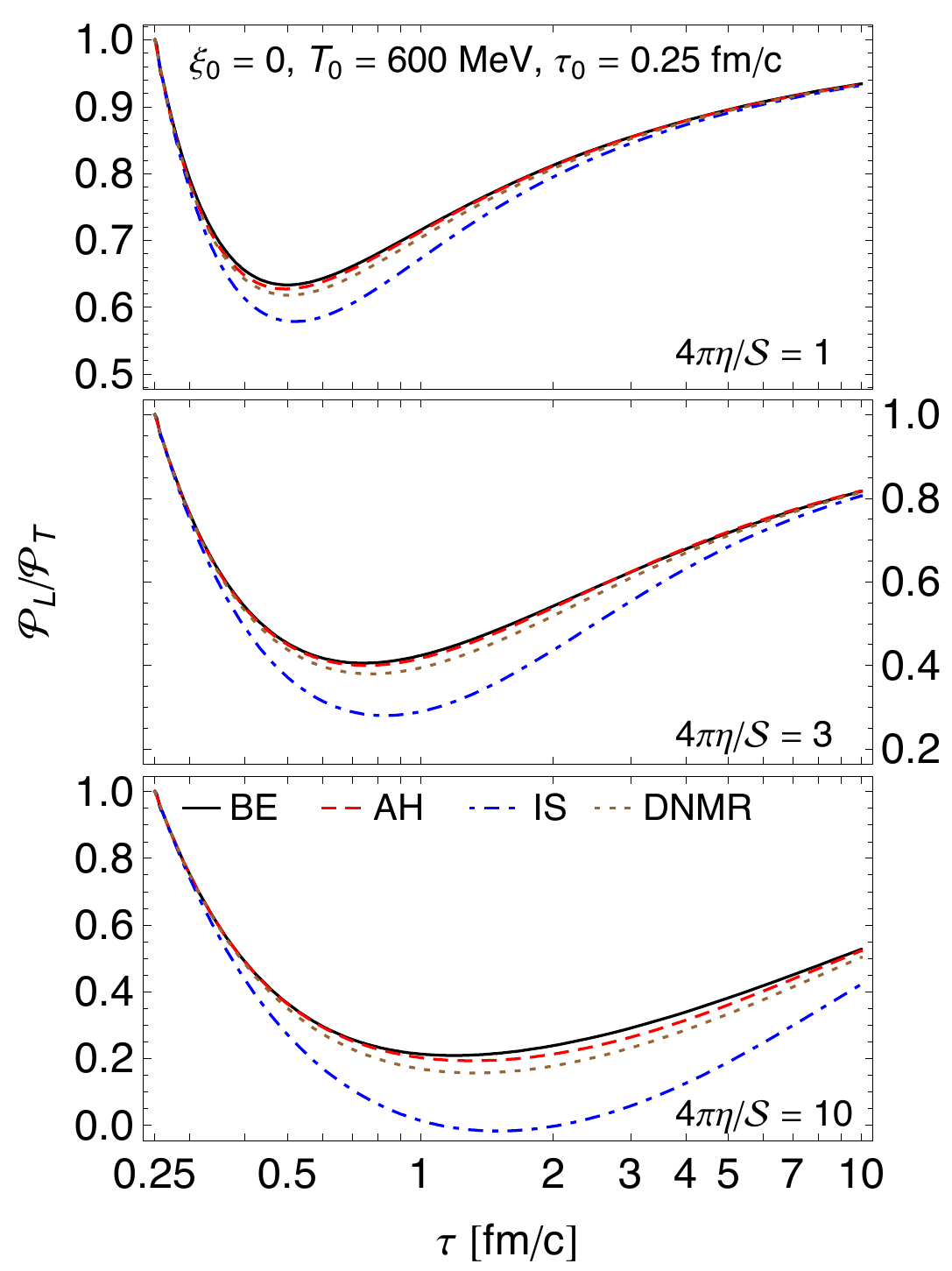}
\includegraphics[width=0.48\linewidth]{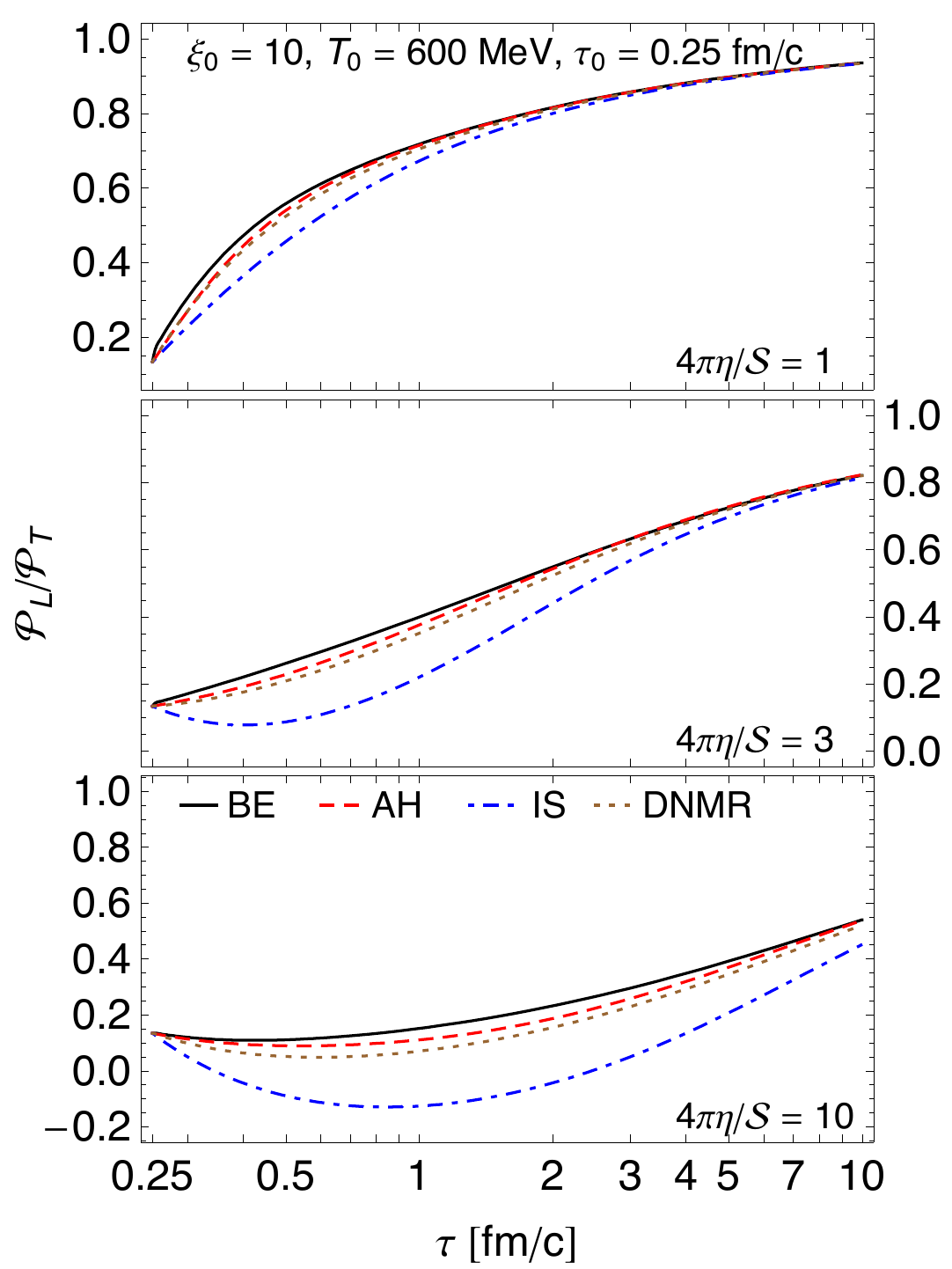}
\end{center}
\caption{(Color online) Pressure anisotropy as a function of proper time assuming $T_0 = 300$ MeV at $\tau_0 =$ 0.25 fm/c for $4 \pi \eta/S =$ 1 (top), 3 (middle), and 10 (bottom).  The left column shows the case $\xi_0 = 0$ and the right column shows the case $\xi_0 = 10$.  The exact solution (black solid), aHydro (AH) approximation (red long-dashed), Israel-Stewart (IS) approximation (blue dot-dashed), and full second order (DNMR) approximation (brown dotted) \cite{Denicol:2012cn} are compared.  Figure adapted from Ref.~\cite{Florkowski:2013lza}.
}
\label{fig:PLoverPT_600}
\end{figure}

\begin{figure}[t]
\begin{center}
\includegraphics[width=0.47\linewidth]{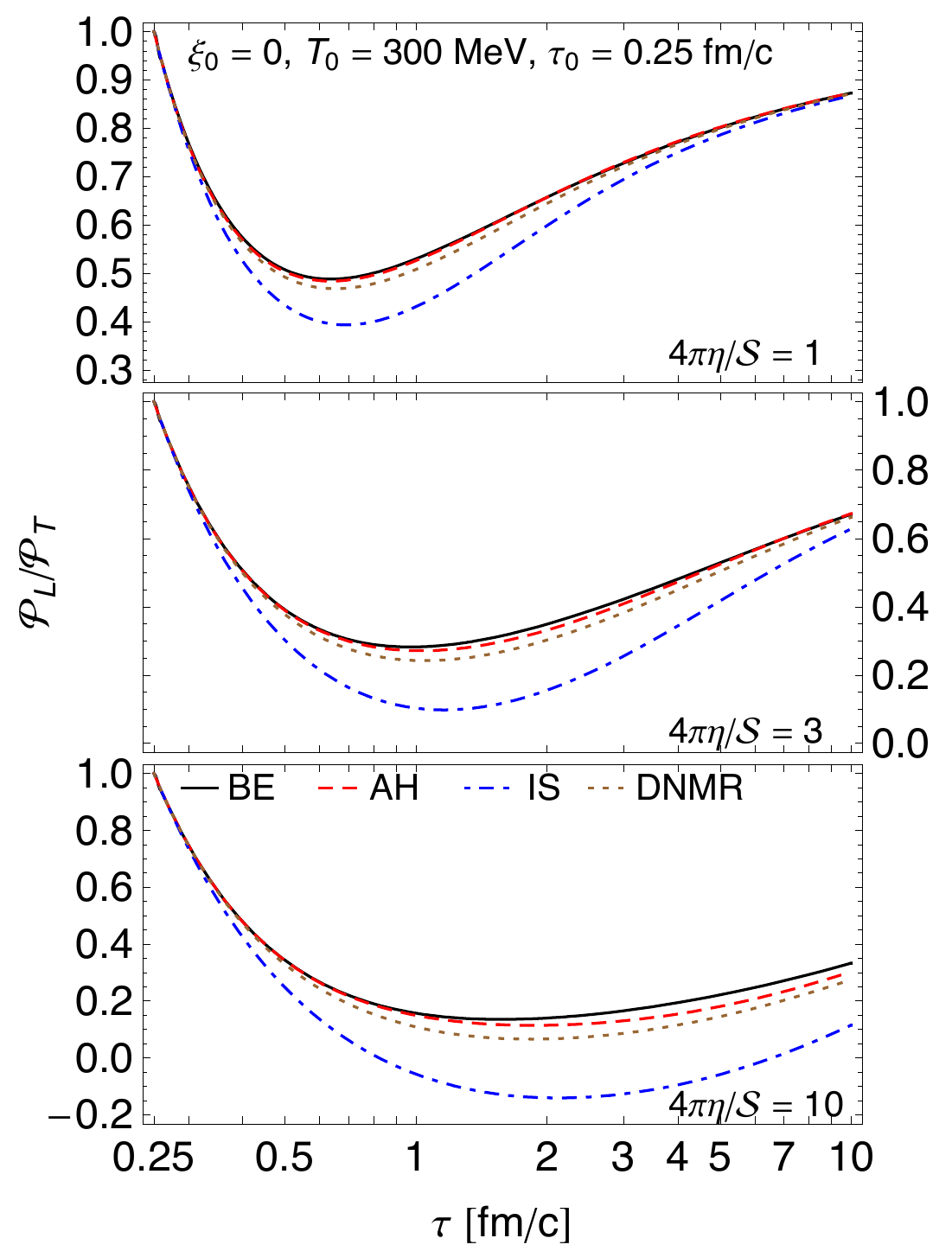}
\includegraphics[width=0.49\linewidth]{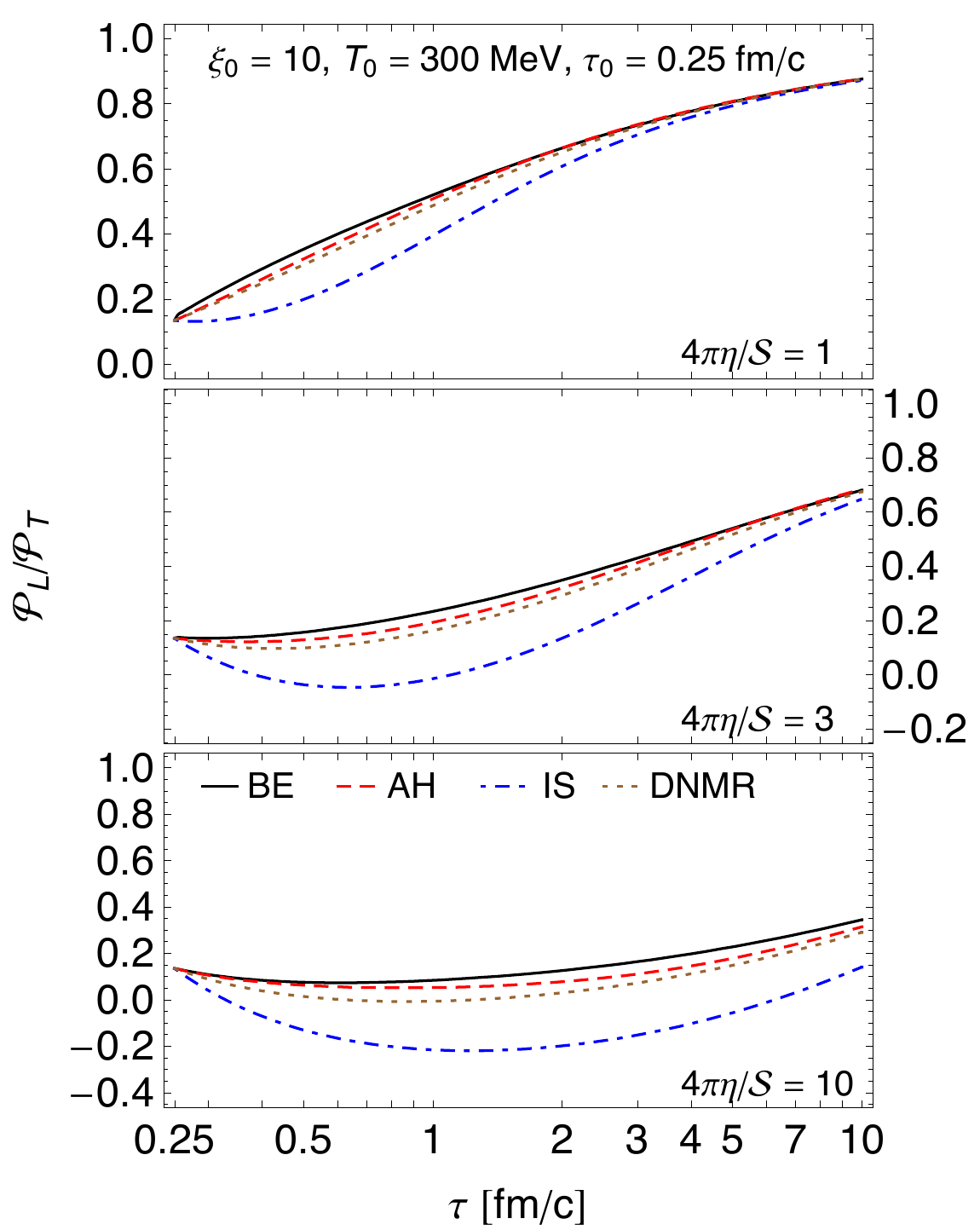}
\end{center}
\caption{(Color online) Pressure anisotropy as a function of proper time assuming $T_0 = 300$ MeV at $\tau_0 =$ 0.25 fm/c for $4 \pi \eta/S =$ 1 (top), 3 (middle), and 10 (bottom).  The left column shows the case $\xi_0 = 0$ and the right column shows the case $\xi_0 = 10$.  Labeling is the same as in Fig.~\ref{fig:PLoverPT_600}.  Figure adapted from Ref.~\cite{Florkowski:2013lza}.}
\label{fig:PLoverPT_300}
\end{figure}

For their results, Florkowski et al. assumed that the initial distribution function was spheroidal in form but for the exact solution they did not restrict the form of the distribution function after this point in time.
In Fig.~\ref{fig:PLoverPT_600} I show the pressure anisotropy as a function of proper time assuming $T_0 = 600$ MeV at $\tau_0 =$ 0.25 fm/c.  The three rows show three different assumed values of the shear viscosity to entropy ratio corresponding to $4 \pi \eta/{\cal S} \in \{1,3,10\}$.  The left column shows the case $\xi_0 = 0$ and the right column shows the case $\xi_0 = 10$.
In the figure the aHydro, Israel-Stewart, and DNMR method \cite{Denicol:2012cn} are compared with the exact solution of the Boltzmann equation.  In Fig.~\ref{fig:PLoverPT_300}, I present the pressure anisotropy subject to the same initial conditions and values of $\eta/{\cal S}$ for an initial effective temperature of $T_0 = 300$ MeV.

As Figs.~\ref{fig:PLoverPT_600} and \ref{fig:PLoverPT_300} demonstrate, the aHydro approximation is always closer to the exact solution than the IS and DNMR approximations.  The IS approximation is the worst approximation to the exact solution in all cases shown and, for the case $4\pi\eta/{\cal S} = 10$, it even predicts a negative longitudinal pressure for the majority of the time shown.  The DNMR approximation represents a significant improvement over the IS approximation; however, we note that, if one increases the shear viscosity to entropy ratio even further, the DNMR approximation also predicts negative longitudinal pressures.  Within the LO aHydro approximation, on the other hand, the pressures are guaranteed to be positive at all times.  
In addition, within LO aHydro, the one-particle distribution function is guaranteed to be positive at all times.

Next, I will review a recent paper that demonstrated that, if one goes to NLO aHydro, the agreement between aHydro and the exact solution presented above becomes extremely good \cite{Bazow:2013ifa}.  Before entering this discussion, I would like to mention that recently it has been demonstrated that the exact solution obtained by Florkowski et al. can be extended \cite{Denicol:2014xca,Denicol:2014tha} to a boost-invariant system that is also expanding transversely subject to ``Gubser flow'' \cite{Gubser:2010ze,Gubser:2010ui}.  This groundbreaking solution will allow practioners to test different dissipative hydrodynamics frameworks in a highly non-trivial case.

\subsection{Next-to-leading-order (NLO) anisotropic hydrodynamics}

Anisotropic hydrodynamics can be extended to NLO by generalizing Eq.~(\ref{eq:ahexp2}) to include arbitrary (but, in principle, small) corrections to a spheroidal LO distribution function~\cite{Bazow:2013ifa}
\begin{equation}
\label{eq14}
f(x,p) = 
f_\mathrm{iso}\left(\frac{\sqrt{p_\mu\Xi^{\mu\nu}(x) p_\nu}}{\Lambda(x)},\frac{\tilde\mu(x)}{\Lambda(x)}\right) 
+ \delta\tilde f(x,p) .
\end{equation}
The parameters $\Lambda$ and $\tilde{\mu}$ are fixed by requiring $\langle E\rangle_{\tilde\delta} = \langle E^2\rangle_{\tilde\delta} = 0$.  To fix the value of the anisotropy parameter $\xi$ one demands that $\delta\tilde{f}$ does not contribute to the pressure anisotropy $\mathcal{P}_T{-}\mathcal{P}_L$. Using Eq.~(\ref{eq14}), one obtains the NLO aHydro decomposition 
\ba
\label{eq15}
j^\mu &=& j^\mu_\mathrm{RS} +\tilde{V}^\mu \, , \\
T^{\mu\nu} &=& T^{\mu\nu}_\mathrm{RS} - \tilde\Pi \Delta^{\mu\nu} + \tilde\pi^{\mu\nu} \, ,
\ea
where
\ba
\tilde\Pi &\equiv& -{\textstyle\frac{1}{3}} \bigl\langle p^{\langle\alpha\rangle} 
p_{\langle\alpha\rangle}\bigr\rangle_{\tilde\delta} \, , \\
\tilde{\pi}^{\mu\nu} &\equiv& \bigl\langle p^{\langle\mu} p^{\nu\rangle}\bigr\rangle_{\tilde\delta} \, , \\
\tilde{V}^\mu &\equiv& \bigl\langle p^{\langle\mu\rangle}\bigr\rangle_{\tilde\delta} \, ,
\ea
with
\begin{equation}
\label{eq2}
\langle O(p)\rangle_{\tilde\delta} \equiv \int \! dP \, O(p)\, \delta\tilde f(x,p) \, .
\end{equation}

The equations above are subject to the constraints $u_\mu \tilde\pi^{\mu\nu}\equiv\tilde\pi^{\mu\nu} u_\nu\equiv(x_\mu x_\nu{+}y_\mu y_\nu{-}2 z_\mu z_\nu)\tilde\pi^{\mu\nu}\equiv{{\tilde\pi}^\mu\!}_\mu\equiv0$. The additional shear stress $\tilde\pi^{\mu\nu}$ arising from $\delta \tilde{f}$ has only 4 degrees of freedom. The strategy of NLO aHydro is to solve the equations of motion for LO aHydro non-perturbatively while coupling the LO equations to additional viscous flows from $\delta\tilde f$.  To close the system, one derives ``perturbative'' second-order equations of motion for $\tilde\Pi$, $\tilde V^\mu$, and $\tilde\pi^{\mu\nu}$. 

\begin{figure}[t]
\begin{center}
\includegraphics[width=0.7\linewidth]{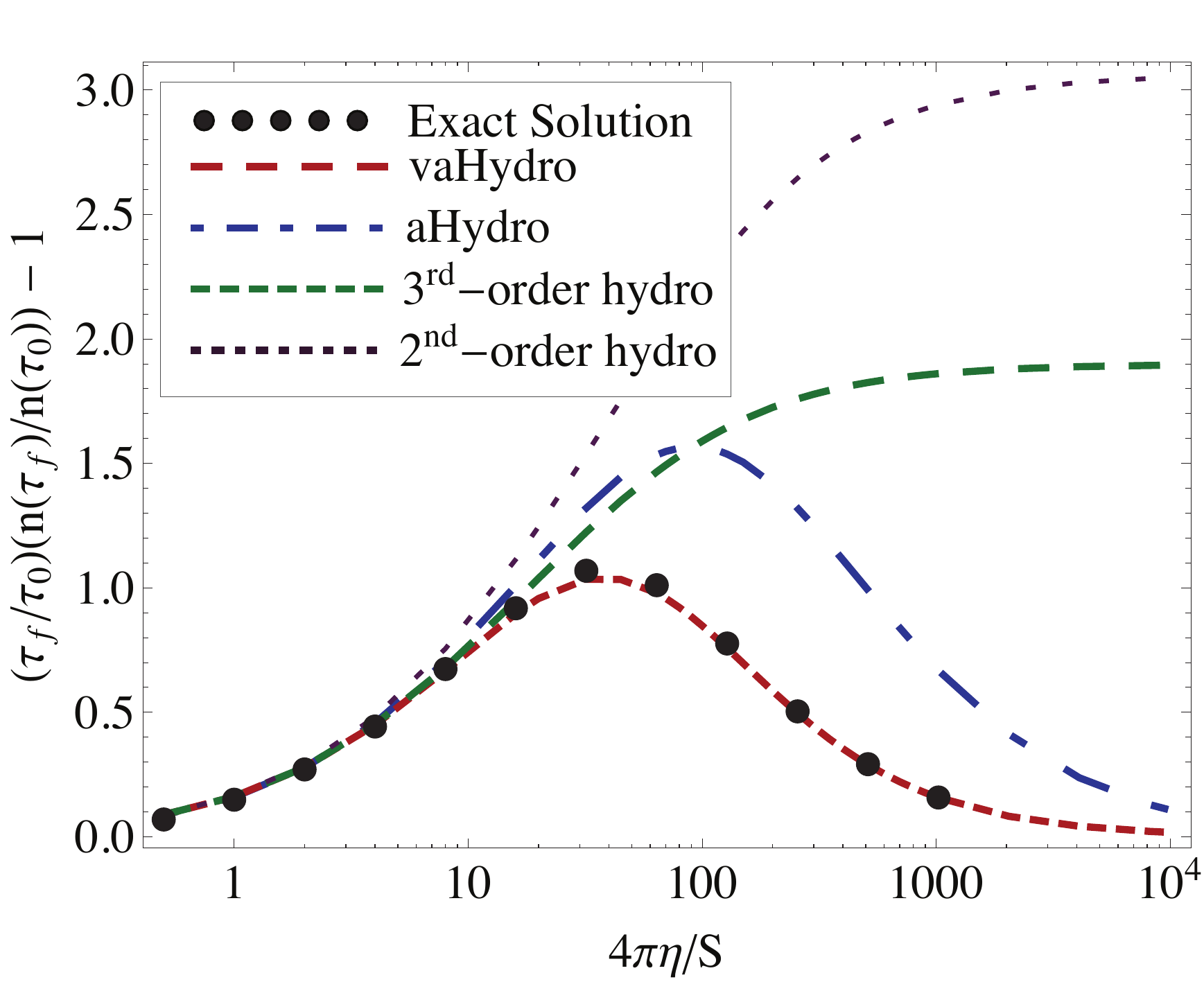}
\end{center}
\caption{(Color online) The particle production measure $(\tau_f n(\tau_f))/(\tau_0 n(\tau_0)) -1$ as a function of $4\pi\eta/s$. The black points, red dashed line, blue dashed-dotted line, green dashed line, and purple dotted line correspond to the exact solution of the Boltzmann equation, {\sc vaHydro}, aHydro, third-order viscous hydrodynamics \cite{Jaiswal:2013vta}, and second-order viscous hydrodynamics \cite{Denicol:2012cn}, respectively. The initial conditions are $T_0 = 600$\,MeV, $\xi_0 = 0$, and $\tilde\pi_0 = 0$ at $\tau_0 = 0.25$\,fm/$c$. The freeze-out temperature was taken to be $T_f = 150$\,MeV.  Figure taken from Ref.~\cite{Bazow:2013ifa}.}
\label{fig:nloahydroeg}
\end{figure}

In Ref.~\cite{Bazow:2013ifa}, the perturbative evolution equations for the dissipative flows $\tilde \Pi,\, \tilde V^\mu$, and $\tilde\pi^{\mu\nu}$ were obtained by generalizing the 14-moment approximation detailed in Ref.~\cite{Denicol:2012cn} to an anisotropic background distribution. The resulting equations are lengthy and I refer the reader to Ref.~\cite{Bazow:2013ifa} for the details. Below I will present the simplified form obtained for 0+1d expansion. I note that $\delta\tilde f$ is much smaller than $\delta f$, particularly at early times, since the largest part of $\delta f$ is already accounted for by the momentum deformation in the LO term of Eq.~(\ref{eq14}). The inverse Reynolds number $\tilde{\mathrm{R}}_\pi^{-1}=\sqrt{\tilde\pi^{\mu\nu}\tilde\pi_{\mu\nu}}/{\cal P}_\mathrm{iso}$ associated with the residual shear stress $\tilde\pi^{\mu\nu}$ is therefore strongly reduced compared to that associated with $\pi^{\mu\nu}$.  This significantly improves the range of applicability of NLO aHydro relative to standard second-order viscous hydrodynamics.

\mysubsubsection{0+1d NLO aHydro}

As mentioned above, for a transversally homogeneous system undergoing boost-invariant longitudinal expansion, the Boltzmann equation can be solved exactly in RTA \cite{Florkowski:2013lya}.  The resulting exact solution can be used to test various macroscopic hydrodynamic approximation schemes. Setting homogeneous initial conditions in $r$ and space-time rapidity $\varsigma$ and zero transverse flow, $\tilde\pi^{\mu\nu}$ reduces to a single non-vanishing component $\tilde\pi$: $\tilde\pi^{\mu\nu}=\mathrm{diag}(0,-\tilde\pi/2,-\tilde\pi/2,\tilde\pi)$ at $z=0$. Using the factorization of the spheroidal energy density, pressures, etc.  presented in the last lecture, one can obtain the following equations of motion for $\dot\xi,\, \dot\Lambda,\,\dot{\tilde\pi}$ \cite{Bazow:2013ifa}
\begin{eqnarray}
&&\frac{\dot\xi}{1{+}\xi}-6\frac{\dot\Lambda}{\Lambda}=
\frac{2}{\tau}+\frac{2}{\tau_{\rm eq}}\left(1-\sqrt{1{+}\xi}\,{\cal R}^{3/4}(\xi)\right) , 
\\
&&{\cal R}'(\xi)\, \dot\xi + 4 {\cal R}(\xi) \frac{\dot\Lambda}{\Lambda} =
- \Bigl({\cal R}(\xi) + {\textstyle\frac{1}{3}} {\cal R}_L(\xi)\Bigr) \frac{1}{\tau} 
+\frac{\tilde\pi}{{\cal E}_\mathrm{iso}(\Lambda)\tau} \, ,
\\
&&\dot{\tilde\pi}=
-\frac{1}{\tau_{\rm eq}}\Bigl[\bigl({\cal R}(\xi){\,-\,}{\cal R}_{\rm L}(\xi)\bigr)P_\mathrm{iso}(\Lambda)+\tilde\pi\Bigr] -\lambda(\xi)\frac{\tilde\pi}{\tau}
\nonumber \\
&& \;\;\; +12\biggl[
   \frac{\dot{\Lambda}}{3\Lambda}\Bigl({\cal R}_{\rm L}(\xi){\,-\,}{\cal R}(\xi)\Bigr)
  +\Bigl(\frac{1{+}\xi}{\tau}-\frac{\dot{\xi}}{2}\Bigr)
  \Bigl({\cal R}^{zzzz}_{-1}(\xi){\,-\,}\frac{1}{3}{\cal R}^{zz}_{1}(\xi)\Bigr)
  \biggr] P_\mathrm{iso}(\Lambda) ,
  \nonumber \\
\end{eqnarray}
where a dot over a symbol indicates a comoving derivative.  The special function $\lambda(\xi)$ and the $\mathcal{R}$-functions appearing above can be found in \cite{Bazow:2013ifa}. The relaxation time $\tau_{\rm eq}$ and the ratio of shear viscosity $\eta$ to entropy density ${\cal S}$, $\eta/{\cal S}$, are related by $\tau_{\rm eq}=5\eta/{\cal S}T= 5\bar\eta/\mathcal{R}^{1/4}(\xi)\Lambda$. In \cite{Bazow:2013ifa} the solution of these equations was compared with the exact solution and various hydrodynamic approximation schemes discussed above plus a 3rd-order viscous hydrodynamic approximation derived in \cite{Jaiswal:2013vta}. As an example of the improvement given by going to NLO, in Fig.~\ref{fig:nloahydroeg}, I show the entropy production (measured by the increase in particle number $\tau n(\tau)$) between the start and the end of the dynamical evolution.  The initial temperature was taken to be  600 MeV and the freeze-out temperature was taken to be 150 MeV. For this extreme 0+1d scenario, where the difference between longitudinal and transverse expansion rates is maximal, NLO aHydro is seen to reproduce the exact solution almost perfectly, dramatically outperforming all other hydrodynamic approximations.

\subsection{Other recent advances}

In this last year, there have been some other important advances that need to be mentioned.  Firstly, I would like to mention the work of Florkowski and Tinti \cite{Tinti:2013vba}.  In this paper, the authors derived dynamical aHydro equations appropriate for describing the spatiotemporal evolution of a 1+1d cylindrically symmetric system.  Instead of using a spheroidal ansatz for the anisotropy tensor, they started from a, more general, ellipsoidal ansatz.  In addition, they demonstrated that, in order to more naturally connect to standard second-order viscous hydrodynamics approaches, it was better to use the second-moment of the Boltzmann equation instead of the zeroth-moment.  

\begin{figure}[t]
\begin{center}
\includegraphics[width=0.75\linewidth]{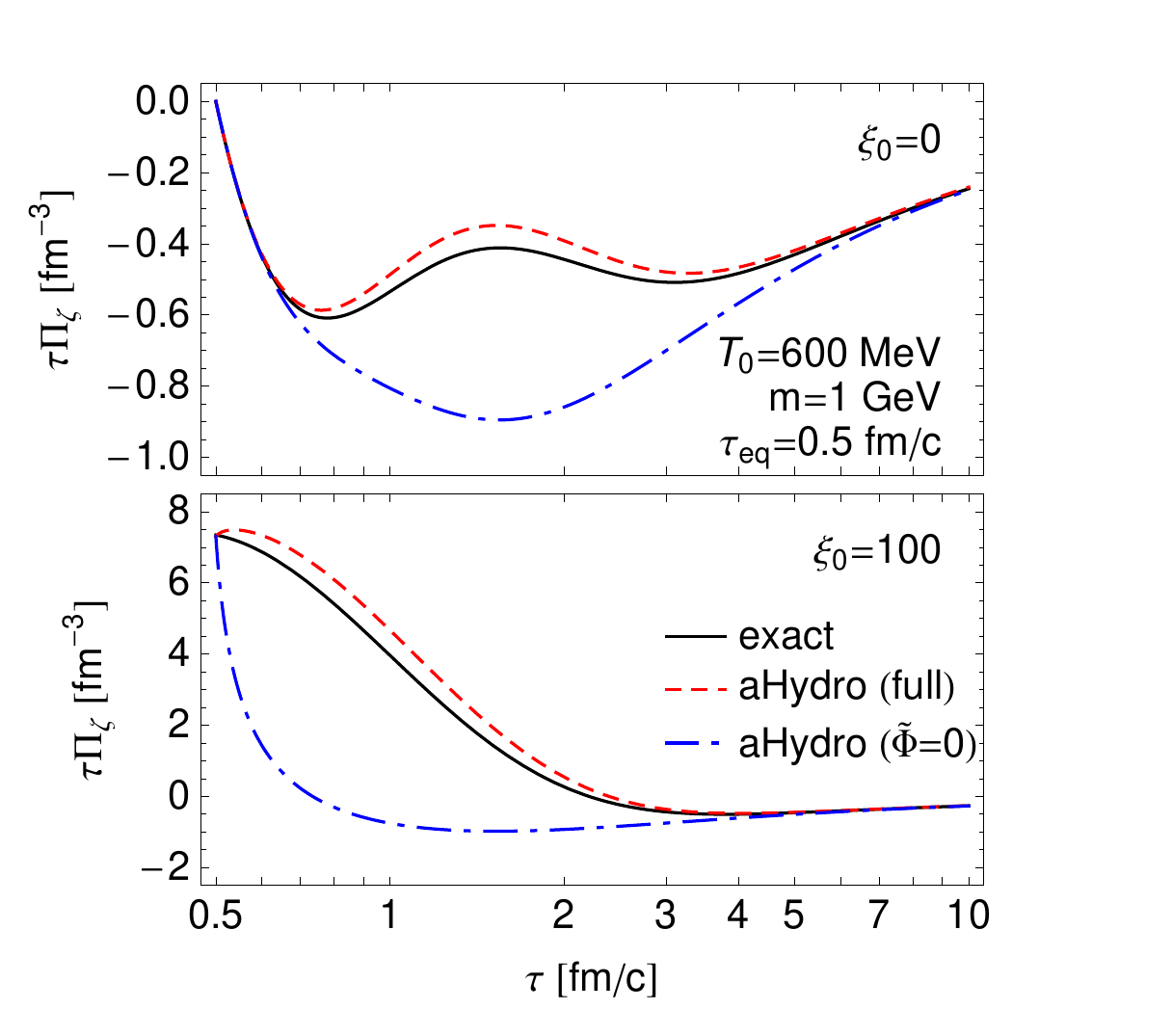}
\end{center}
\caption{(Color online) Proper-time evolution of the bulk pressure.  The three lines correspond to the exact solution of the Boltzmann equation \cite{Florkowski:2014sfa} (black solid line), the full aHydro equations including the bulk degree of freedom (red dashed line), and the aHydro equations with the ellipsoidal bulk degree of freedom set to zero (blue dot-dashed line).  For both panels we used $m=$ 1 GeV, $\tau_0$ = 0.5 fm/c, $\tau_{\rm eq}$ = 0.5 fm/c, and $T_0$ = 600 MeV.  In the top panel we fixed the initial spheroidal anisotropy parameter $\xi_0=0$ and in the bottom panel we chose $\xi_0 = 100$.  Figure taken from Ref.~\cite{Nopoush:2014pfa}.}
\label{fig:bulk_1000}
\end{figure}

In a subsequent paper by Florkowski et al \cite{Florkowski:2014sfa}, it was shown that using the equation obtained from the second moment yielded better agreement with exact solutions of the RTA Boltzmann equation for massive particles obtained in Ref.~\cite{Florkowski:2014sfa}.  However, Ref.~\cite{Florkowski:2014sfa} found that, although aHydro worked better than Israel-Stewart theory, the evolution of the bulk pressure was still rather poorly described compared to the description of the pressure anisotropy.  To address this, in a subsequent paper, Nopoush et al. generalized the formalism of Tinti and Florkowski to include a explicit degree of freedom associated with the bulk pressure \cite{Nopoush:2014pfa}.  Ref.~\cite{Nopoush:2014pfa} also showed that, when including the bulk degree of freedom, the additional equation of motion necessary could be provided by the zeroth moment.  Comparisons of numerical results obtained with the dynamical equations obtained in Ref.~\cite{Nopoush:2014pfa} and the exact solution obtained in Ref.~\cite{Florkowski:2014sfa} showed that the inclusion of an explicit bulk degree of freedom dramatically improved the agreement of aHydro with the exact solution.  As a demonstration of this improvement, in Fig.~\ref{fig:bulk_1000}, I show the proper-time evolution of the bulk pressure for the case $m=$ 1 GeV, $\tau_0$ = 0.5 fm/c, $\tau_{\rm eq}$ = 0.5 fm/c, and $T_0$ = 600 MeV.  The figure shows the aHydro result obtained with and without the explicit bulk degree of freedom ($\Phi$) included.  As can be seen from this figure, inclusion of the bulk degree of freedom dramatically improves agreement with the exact RTA Boltzmann equation solution.

\section{Conclusions}

In this writeup, I have attempted to convey the content of the three lectures I gave at the LIV Cracow School of Theoretical Physics.  There has been a lot of progress in the area of dissipative hydrodynamics, including anisotropic hydrodynamics, in recent years.  The recent extension of aHydro to NLO provides a complete second-order treatment which takes into account QGP momentum-space anisotropies from the outset and, as a result, yields a superior approximation scheme.  Future developments will include implementation of numerical codes including anisotropic freeze out.  Since the expansion around a locally anisotropic momentum distribution results in smaller deviations $\delta\!\tilde{f}$ of the distribution function from the leading-order ansatz, the NLO aHydro framework should yield results that are quantitatively more reliable, particularly when it comes to the early stages of QGP hydrodynamical evolution and near the transverse edges of the overlap region where the system is approximately free streaming.  As mentioned previously, another important recent development has been the development of leading-order ellipsoidal anisotropic hydrodynamics \cite{Tinti:2013vba}.  Before wrapping up, I would like emphasize that, now that it is widely accepted that the QGP is momentum-space anisotropic, many fundamental QGP processes should be carefully reconsidered.  Finally, I will note that a spin-off of these studies has been the development a new exact solution of the RTA Boltzmann equation which includes simultaneous transverse and longitudinal expansion \cite{Denicol:2014xca,Denicol:2014tha}.  This development offers the possibility to quantitatively assess the accuracy of different hydrodynamic approaches.

\section*{Acknowledgments}

I thank the organizers of the LIV Cracow School of Theoretical Physics for inviting me to present these lectures.  I also thank the participants of the summer school for humoring me when I attempted to play fiddle with the local Polish highlanders.  Finally, I thank D.~Bazow, U.~Heinz, E.~Maksymiuk, M.~Martinez, M.~Nopoush, R.~Ryblewski, and L.~Tinti for their collaboration on anisotropic hydrodynamics.  Support for this work was provided in the framework of the JET Collaboration by U.S. DOE Award No.~\rm{DE-AC0205CH11231}.

\bibliographystyle{utphys}
\bibliography{strickland}

\end{document}